\documentclass[12pt,preprint]{aastex}

\shorttitle{Non-spherical quasi-relaxed stellar systems}
\shortauthors{Bertin and Varri}

\bibpunct[; ]{(}{)}{;}{a}{}{;}

\begin{document}

\title{The construction of non-spherical models of quasi-relaxed stellar systems}

\author{G. Bertin and A.L. Varri}
\affil{Universit\`{a} degli Studi di Milano, Dipartimento di Fisica, via Celoria 16, I-20133 Milano, Italy; anna.varri@studenti.unimi.it}

\begin{abstract}

Spherical models of collisionless but quasi-relaxed stellar systems 
have long been studied as a natural framework for the description of 
globular clusters. Here we consider the construction of self-consistent 
models under the same physical conditions, but including explicitly 
the ingredients that lead to departures from spherical symmetry.
In particular, we focus on the effects of the tidal field associated 
with the hosting galaxy. We then take a stellar system on a circular 
orbit inside a galaxy represented as a ``frozen" external field. The 
equilibrium distribution function is obtained from the one describing 
the spherical case by replacing the energy integral with the relevant 
Jacobi integral in the presence of the external tidal field. Then the 
construction of the model requires the investigation of a singular 
perturbation problem for an elliptic partial differential equation with 
a free boundary, for which we provide a method of solution to any desired
order, with explicit solutions to two orders. We outline the relevant 
parameter space, thus opening the way to a systematic study of the
properties of a two-parameter family of physically justified
non-spherical models of quasi-relaxed stellar systems. The general
method developed here can also be used to construct models for which 
the non-spherical shape is due to internal rotation. Eventually, the 
models will be a useful tool to investigate whether the shapes of
globular clusters are primarily determined by internal rotation,
by external tides, or by pressure anisotropy.

\end{abstract}


\section{Introduction}

Large stellar systems can be studied as collisionless systems, by
means of a one-star distribution function obeying the combined set
of the collisionless Boltzmann equation and the Poisson equation,
under the action of the self-consistent mean potential. For elliptical 
galaxies the relevant two-star relaxation times do actually exceed 
their age; an imprint of partial relaxation may be left at the time 
of their formation \citetext{if we refer to a picture of formation 
via incomplete violent relaxation; \citealp{Lyn67}, \citealp{Alb82}}, 
but otherwise they should be thought of as truly collisionless systems, 
generally characterized by an anisotropic pressure tensor. In turn, 
for globular clusters the relevant relaxation times are typically 
shorter than their age, so that we may argue that for many of them 
the two-star relaxation processes have had enough time to act and 
to bring them close to a thermodynamically relaxed state, with their 
distribution function close to a Maxwellian. This line of arguments 
has led to the development of well-known dynamical models for globular 
clusters \citep{Kin65,Kin66}.

King models are based on a quasi-Maxwellian isotropic distribution
function $f_K(E)$ in which a truncation prescription, continuous
in phase space, is set heuristically to incorporate the presence
of external tides; but otherwise, they are fully self-consistent
(i.e., no external fields are actually considered) and perfectly
spherical. Empirically, the simplification of spherical symmetry
is encouraged by the fact that in general globular clusters have
round appearance. Indeed, as a zero-th order description, these
models have had remarkable success in applications to observed
globular clusters \citep[e.g., see][and references therein]{Spi87,
DjoMey94,McLMar05}.
In recent years, great progress has been made in the acquisition of
detailed quantitative information about the structure of these stellar
systems, especially in relation to the measurement of the proper
motions of thousands of individual stars \citep[see][]{Lee00,McL06}, 
with the possibility of getting a direct
5-dimensional view of their phase space. Such progress calls for
renewed efforts on the side of modeling. More general models would
allow us to address the issue of the origin of the observed
departures from spherical symmetry. In fact, it remains to be
established which physical ingredient among rotation, pressure
anisotropy, and tides is the primary cause of the flattening of
globular clusters \citep[e.g., see][and references therein]{Kin61,
FreFal82,Gey83,WhiSha87,DavPru90,HanRyd94,Ryd96,Goo97,Ber08} 

As in the case of the study of elliptical galaxies \citep[e.g., see]
[and references therein]{BerSti93}, different approaches can be 
taken to the construction of models.
Broadly speaking, two complementary paths can be followed. In the
first, ``descriptive" approach, under suitable geometrical (on the
intrinsic shape) and dynamical (e.g., on the absence or presence
of dark matter) hypotheses, the available data for an individual
stellar system are imposed as constraints to derive the internal
orbital structure (distribution function) most likely to
correspond to the observations. This approach is often carried out
in terms of codes that generalize a method introduced by
\citet{Sch79}; for an application to the globular cluster
$\omega$ Cen, see \citet{Ven06}. In the second, ``predictive" 
approach, one proposes a formation/evolution scenario in order to 
identify a physically justified distribution function for a wide 
class of objects, and then proceeds to investigate, by comparison 
with observations of several individual objects, whether the data 
support the general physical picture that has been proposed. Indeed, 
King models belong to this latter approach.

The purpose of this paper is to extend the description of
quasi-relaxed stellar systems, so far basically limited to the
spherical King models, to the non-spherical case. There are at
least three different ways of extending spherical isotropic models
of quasi-relaxed stellar systems (such as King models), by
modifying the distribution function so as to include: (i) the
explicit presence of a non-spherical tidal field; (ii) the
presence of internal rotation; (iii) the presence of some pressure
anisotropy. As noted earlier in this Introduction, these
correspond to the physical ingredients that, separately, may be
thought to be at the origin of the observed non-spherical shapes.
We will thus focus on the construction of physically justified
models, as an extension of the King models, in the presence of
external tides and, briefly, on the extension of the models to the
presence of rigid internal rotation. A first-order analysis of the
triaxial tidal problem addressed in this paper was carried out by
\citet{HegRam95} and the effect of a ``frozen'' tidal field on 
(initially) spherical King models was studied by \citet{Wei93} 
using N-body simulations. 
For the axisymmetric problem associated with the presence of internal 
rotation, a first-order analysis of a simply truncated Maxwellian 
distribution perturbed by a rigid rotation was given by \citet{KorAna71}; 
different models where differential rotation is considered were 
proposed by \citet{PreTom70} and by \citet{Wil75}, also in view of
extensions to the presence of pressure anisotropy, which goes beyond
the scope of this paper. Models that represent a direct generalization
of the King family to the case with differential rotation have also 
been examined, with particular attention to their thermodynamic 
properties \citep{LagLon96,LonLag96}. In principle, the method of 
solution that we will present below can deal with the extension of 
other spherical isotropic models with finite size that are not 
of the King form \citetext{e.g., see \citealp{WooDic62,Dav77}; 
see also the interesting suggestion by \citealp{Mad96}}.

The study of self-consistent collisionless equilibrium models has
a long tradition not only in stellar dynamics, but also in plasma
physics \citep[e.g., see][]{Har62,AttPeg99}. We note
that in both research areas a study in the presence of external
fields, especially when the external field is bound to break the
natural symmetry associated with the one-component problem, is
only rarely considered.

The paper is organized as follows. Section 2 introduces the
reference physical model, in which a globular cluster is imagined
to move on a circular orbit inside a host galaxy treated as a
frozen background field; the modified distribution function for
such a cluster is then identified and the relevant parameter space
defined. In Sect.~3 we set the mathematical problem associated
with the construction of the related self-consistent models. For
models generated by the spherical $f_K(E)$, Sect.~4 and Appendix 
A give the complete solution in terms of matched asymptotic 
expansions. Alternative methods of solutions are briefly discussed 
in Sect.~5. The concluding Section 6 gives a summary of the paper, 
with a short discussion of the results obtained. In Appendix B we 
show how the method developed in this paper can be applied to 
construct quasi-relaxed models flattened by rotation in the absence 
of external tides. In Appendix C we show how the method can be 
applied to other isotropic truncated models, different from King 
models. 

Technically, the mathematical problem of a singular perturbation 
with a free boundary that is faced here is very similar to the 
problem noted in the theory of rotating stars, starting with Milne 
\citep[see][]{Tas78,Mil23,Cha33,Kro42,ChaLeb62,MonRox65}. The 
problem was initially dealt with inadequate tools; a satisfactory 
solution of the singular perturbation problem was obtained only 
later, by \citet{Smi75,Smi76}.

\section{The physical model}

\subsection{The tidal potential}

As a reference case, we consider an idealized model in which the
center of mass of a globular cluster is imagined to move on a
circular orbit of radius $R_0$, characterized by orbital frequency
$\Omega$, inside a host galaxy. For simplicity, we focus on the
motion of the stars inside the globular cluster and model the
galaxy, taken to have very large mass, by means of a {\it frozen}
gravitational field (which we will call the galactic field,
described by the potential $\Phi_G$), with a given overall
symmetry. This choice makes us ignore interesting effects that are
generally present in the full interaction between a ``satellite"
and a galaxy; in a sense, we are taking a complementary view of an
extremely complex dynamical situation, with respect to other
investigations, such as those that lead to a discussion of the
mechanism of dynamical friction \citetext{in which the globular 
cluster or satellite is modeled as a rigid body and the stars of 
the galaxy are taken as the ``live" component; see \citealp{BonAlb87}; 
\citealp{Ber03}; \citealp{AreBer07}; and references therein}. 
Therefore, we will be initially following the picture of a 
{\it restricted three-body problem}, with one important
difference, that the ``secondary" is not treated as a point mass
but as a ``live" stellar system, described by the cluster
mean-field potential $\Phi_C$. In this extremely simple orbital
choice for the cluster center of mass, in the corotating frame the
associated tidal field is time-independent and so we can proceed
to the construction of a stationary dynamical model.

We consider the galactic potential $\Phi_G$ to be spherically
symmetric, i.e. $\Phi_G = \Phi_G (R)$, with $R= \sqrt{X^2
+Y^2+Z^2}$, in terms of a standard set of Cartesian coordinates
$(X,Y,Z)$, so that $\Omega^2 = (d\Phi_G (R)/dR)_{R_0}/R_0$. Let
$(X,Y)$ be the orbit plane of the center of mass of the cluster.
We then introduce a local rotating frame of reference, so that the
position vector is given by ${\bf r}=(x,y,z)$, with origin in the
center of mass of the cluster and for which the x-axis points away
from the center of the galactic field, the y-axis follows the
direction of the cluster rotation in its orbit around the galaxy,
and the z-axis is perpendicular to the orbit plane (according to
the right-hand rule). In such rotating local frame, the relevant
Lagrangian, describing the motion of a star belonging to the
cluster, is \citetext{cf. \citealp{Cha42}, Eq.~[5.510]}:
\begin{equation}\label{lagran}
\mathcal{L}=\frac{1}{2}\{\dot{x}^2+\dot{y}^2+\dot{z}^2 +\Omega^2[(R_0+x)^2+y^2]
+2\Omega(R_0+x)\dot{y}-2\Omega\dot{x}y\}-\Phi_G(R)-\Phi_C(x,y,z)~,
\end{equation}
where $R =\sqrt{(R_0+x)^2+y^2+z^2}$ and the terms responsible for 
centrifugal and Coriolis forces are explicitly displayed.

If we suppose that the size of the cluster is small compared to
$R_0$, we can adequately represent the galactic field by its
linear approximation with respect to the local coordinates (the
so-called ``tidal approximation"). The corresponding equations of
the motion for a single star in the rotating local frame are:
\begin{eqnarray}
&&\displaystyle\ddot{x}-2\Omega \dot{y}-(4\Omega^2-\kappa^2)x=
-\frac{\partial \Phi_C}{\partial x}~, \label{eqmotion1} \\
&&\displaystyle\ddot{y}+2\Omega \dot{x}=-\frac{\partial
\Phi_C}{\partial y}~, \label{eqmotion2} \\
&&\displaystyle\ddot{z}+\Omega^2z=-\frac{\partial \Phi_C}{\partial z}~,\label{eqmotion3}
\end{eqnarray}
where $\kappa$ is the epicyclic frequency at $R_0$, given by 
$\kappa^2 = 3\Omega^2 + (d^2\Phi_G/dR^2)_{R_0}$. Note that the 
assumed symmetry for $\Phi_G$ introduces a cancellation between the 
kinematic term $y \Omega^2$ and the gradient of the galactic 
potential $\partial \Phi_G/ \partial y$ and makes the vertical 
acceleration $ - \partial \Phi_G/ \partial z$ approximately equal 
to $-z\Omega^2$.

These equations admit an energy (isolating) integral of the
motion, known as the Jacobi integral:
\begin{equation}\label{Jacobi}
H = \frac{1}{2}(\dot{x}^2+\dot{y}^2+\dot{z}^2)+\Phi_T+\Phi_C~,
\end{equation}
where
\begin{equation}\label{tidalpot}
\Phi_T=\frac{1}{2}\Omega^2\left(z^2-\nu x^2\right)
\end{equation}
is the tidal potential. Here $\nu\equiv 4 -\kappa^2/\Omega^2$
is a generally positive dimensionless coefficient.

Thus, at the level of single star orbits, we note that, in
general, the tidal potential leads to a compression in the
z-direction, a stretching in the x-direction, and leaves the
y-direction untouched. The tidal potential is static, breaks the
spherical symmetry, but is characterized by reflection symmetry
with respect to the three natural coordinate planes; strictly
speaking, the symmetry with respect to $(y,z)$ is applicable only
in the limit of an infinitely massive host galaxy \citep[see][]{Spi87}. 
In turn, we will see that the geometry of the tidal
potential induces a non-spherical distortion of the cluster shape
collectively, in particular an elongation along the x-axis and a
compression along the z-axis. In practice, the numerical
coefficient $\nu$ that defines quantitatively the induced
distortion depends on the galactic potential. We recall that we
have $\nu = 3$ for a Keplerian potential, $\nu = 2$ for a
logarithmic potential, while for a Plummer model the dimensionless
coefficient depends on the location of the circular orbit with
respect to the model scale radius $b$, with $\nu(R_0)=3 R_0^2/
(b^2+R_0^2)$ \citetext{for a definition of the Plummer model see, e.g.,
\citealp{Ber00}}.

Different assumptions on the geometry of the galactic field can be
treated with tools similar to those developed here, leading to a
similar structure of the equations of the motion, with a slight
modification of the tidal field. In particular, for an
axisymmetric galactic field, the tidal potential differs from the
one obtained here only by the z-term \citep{Cha42,HegHut03}. This 
case is often considered, for example by
referring to a globular cluster in circular orbit on the
(axisymmetric) disk of our Galaxy \citep[see][]{HegRam95,Ern08}, 
for which $\Phi_T$ is then formulated in terms of the Oort constants.

In the physical model outlined in this Section, the typical
dynamical time associated with the star orbits inside the cluster
is much smaller than the (external) orbital time associated with
$\Omega$. Therefore, in an asymptotic sense, the equilibrium
configurations that we will construct in the rest of the paper can
actually be generalized, with due qualifications, to more general
orbits of the cluster inside a galaxy, provided we interpret the
results that we are going to obtain as applicable only to a small
piece of the cluster orbit.

\subsection{The distribution function}

As outlined in the Introduction, we wish to extend the description
of quasi-relaxed stellar systems (so far basically limited to
spherical models associated with distribution functions $f =
f(E)$, dependent only on the single-star specific energy $E =
v^2/2 + \Phi_C$) to the non spherical case, by including the
presence of a non-spherical tidal field explicitly. Given the
success of the spherical King models in the study of globular
clusters, we will focus on the extension of models based on
$f_K(E)$, which is defined as a ``lowered" Maxwellian, continuous
in phase space, with an energy cut-off which implies the existence
of a boundary at the truncation radius $r_{tr}$.

Therefore, we will consider (partially) self-consistent models
characterized by the distribution function:
\begin{equation}\label{fK}
f_K(H)= A [\exp(-aH)- \exp(-aH_0)]
\end{equation}
\noindent if $H \le H_0$ and $f_K(H) = 0$ otherwise, in terms of
the Jacobi integral defined by Eq.~(\ref{Jacobi}). Here $H_0$ is
the cut-off value for the Jacobi integral, while $A$ and $a$ are
positive constants.

In velocity space, the inequality $H \le H_0$ identifies a
spherical region given by $0 \le v^2 \le 2 \psi({\bf r})/a$, where:
\begin{equation}\label{psi}
\psi({\bf r})=a\{H_0-[\Phi_C({\bf r})+\Phi_T(x,z)]\}
\end{equation}
is the dimensionless escape energy. Therefore, the boundary of the 
cluster is defined as the relevant zero-velocity surface by the 
condition $\psi({\bf r}) = 0$ and is given only implicitly by an 
equipotential ({\it Hill}) surface for the total potential $\Phi_C + 
\Phi_T$; in fact, its geometry depends on the properties of the tidal 
potential (of known characteristics; see Eq.~[\ref{tidalpot}]) and 
of the cluster potential (unknown {\it a priori}, to be determined 
as the solution of the associated Poisson equation).

The value of the cut-off potential $H_0$ should be chosen in such
a way that the surface that defines the boundary is closed. The
last (i.e., outermost) closed Hill surface is a {\em critical}
surface, because it contains two saddle points that represent the
Lagrangian points of the restricted three-body problem outlined in
the previous subsection. From
Eqs.~(\ref{eqmotion1})-(\ref{eqmotion3}), we see that such
Lagrangian points are located symmetrically with respect to the
origin of the local frame of reference and lie on the x-axis.
Their distance from the origin is called the {\em tidal radius},
which we denote by $r_T$, and can be determined from the condition:
\begin{equation}\label{tidalradius}
\frac{\partial \psi}{\partial x}(r_T,0,0)=0~.
\end{equation}
If, as a zero-th order approximation, we use a simple Keplerian 
potential for the cluster potential $\Phi_C$, we recover the 
classical expression (e.g., see Spitzer 1987):
\begin{equation}
r_T^{(0)} = \left(\frac{G M}{\Omega^2 \nu} \right)^{1/3}~,
\end{equation}
where $M$ is the total mass of the cluster.

As for the spherical King model, the density profile associated
with the distribution function (\ref{fK}) is given by:
\begin{equation}\label{rho}
\rho ( \psi ) = \hat{A}  e^ \psi \gamma \left({5 \over 2}, \psi
\right) \equiv \hat{A}\hat{\rho}(\psi)~,
\end{equation}
where $\hat{A} = {8 \pi A \sqrt{2}e^{-a H_0} /(3 a^{3/2})}$. We 
recall that the incomplete gamma function has non-negative real 
value only in correspondence to a non-negative argument. In the 
following, we will denote the central density of the cluster by 
$\rho_0 = \hat{A}\hat{\rho}(\Psi)$, where $\Psi\equiv \psi({\bf 0})$ 
is the depth of the central potential well.

\subsection{The parameter space}\label{parspa}

The models defined by $f_K(H)$ are characterized by two physical
scales (e.g., the two free constants $A$ and $a$, or,
correspondingly, the total mass $M$ and the central density
$\rho_0$ of the cluster) and two dimensionless parameters. The
first dimensionless parameter can be defined, as in the spherical
King models, to measure the {\it concentration} of the cluster. We
can thus consider the quantity $\Psi$, introduced at the end of
the previous subsection, or we may refer to the commonly used
concentration parameter:
\begin{equation}\label{concentr}
C = \log(r_{tr}/r_0)~,
\end{equation}
where $r_0 = \sqrt{9/(4 \pi G \rho_0 a)}$ is a scale length related 
to the size of the core and $r_{tr}$ is the truncation radius of 
the spherical King model associated with the same value of the 
central potential well $\Psi$ \citetext{the relation between $C$ 
and $\Psi$ is one-to-one; e.g., see \citealp{Ber00}}.

The second dimensionless parameter characterizes the strength of
the (external) tidal field:
\begin{equation}\label{tidalp}
\epsilon \equiv \frac{\Omega^2}{4 \pi G \rho_0}~.
\end{equation}
The definition arises naturally from the dimensionless formulation 
of the Poisson equation that describes the (partially)
self-consistent problem (to be addressed in the next Section).

In principle, for a given choice of the dimensional scales ($A$
and $a$) the truncation radius or the concentration parameter of a
spherical King model can be set arbitrarily. In practice, the
physical motivation of the models suggests that the truncation
radius $r_{tr}$ should be taken to be of the order of (and not
exceed) the tidal radius $r_T$, introduced in the previous
subsection (see Eq.~[\ref{tidalradius}]). We may thus define an
{\em extension parameter}, as the ratio between the truncation
radius of the corresponding spherical model and the tidal radius
$r_T$:
\begin{equation}\label{delta}
\delta \equiv \frac{r_{tr}}{r_T}~.
\end{equation}
For a given value of the central potential well $\Psi$,
there exists a (maximum) critical value for the tidal strength
parameter, which we will denote by $\epsilon_{cr}$, corresponding
to the maximum value for the extension parameter $\delta_{cr}$,
which can be found by solving the system:
\begin{equation}
\left\{
\begin{array}{ll}\label{sistcrit}
\displaystyle \frac{\partial \psi}{\partial
x}(r_T,0,0;\epsilon_{cr})=0  \vspace{.1cm}\\ \displaystyle
\psi(r_T,0,0;\epsilon_{cr})=0~.
\end{array}
\right.
\end{equation}
From this system, if we use the zero-th order Keplerian approximation 
for $\Phi_C$, we find that $\delta_{cr}^{(0)} = 2/3$ \citep[see][]{Spi87}.

For our two-parameter family of models we thus expect two tidal
regimes to exist. For models characterized by the pairs $(\Psi,
\epsilon)$ near the critical condition $\delta \approx
\delta_{cr}$ the tidal distortion should be maximal, while for
models with pairs well below criticality only small departures
from spherical symmetry should occur. A thorough exploration of
the parameter space will be carried out in a separate paper (Varri
\& Bertin, in preparation). In closing, we note that the models
proposed and studied by \citet{HegRam95} correspond to
the pairs in parameter space that we have called critical.

\section{The mathematical problem}

The (partially) self-consistent models associated with the
distribution function defined by Eq.~(\ref{fK}) are constructed by
solving the relevant Poisson equation. In terms of the
dimensionless escape energy $\psi$, given by Eq.~(\ref{psi}), the
Poisson equation (for $\psi \ge 0$) can be written as:
\begin{equation}
\nabla^2 (\psi + a \Phi_T) = -\frac{9}{r_0^2}\frac{\rho}{\rho_0} =
- \frac{9}{r_0^2}\frac{\hat{\rho}(\psi)}{\hat{\rho}(\Psi)}~,
\end{equation}
where $r_0$ is the scale length introduced in Sect.~\ref{parspa} 
(see Eq.~[\ref{concentr}]). We then rescale the coordinates and 
introduce the dimensionless position vector $\hat{{\bf r}} = {\bf r}
/r_0$, so that $\hat{\nabla}^2=r_0^2\nabla^2$ and $a \Phi_T \equiv 
\epsilon T=9 \epsilon (\hat{z}^2- \nu\hat{x}^2)/2$, where we have 
made use of the tidal parameter introduced in Eq.~(\ref{tidalp}). 
Therefore, the Poisson equation, for $\psi \ge 0$, can be written 
in dimensionless form as:
\begin{equation}\label{Poisson}
\hat{\nabla}^2\psi=-9\left[\frac{\hat{\rho}(\psi)}{\hat{\rho}(\Psi)}
+\epsilon(1-\nu)\right]~,
\end{equation}
while for negative values of $\psi$ we should refer to:
\begin{equation}\label{vacuum}
\hat{\nabla}^2 \psi=-9\epsilon (1-\nu)~,
\end{equation}
i.e. the Laplace equation $\hat{\nabla}^2(a\Phi_C)=0$. 

The mathematical problem is completed by specifying the
appropriate boundary conditions. As for the spherical King models,
we require regularity of the solution at the origin
\begin{eqnarray}
\psi({\bf 0})=\Psi~, \label{cond1}\\ 
\hat{\nabla} \psi({\bf 0})={\bf0}~, \label{cond2}
\end{eqnarray}
and, at large radii:
\begin{equation}
\psi + \epsilon T \rightarrow a H_0 \label{cond3}~,
\end{equation}
which corresponds to $a\Phi_C \rightarrow 0$.

Poisson and Laplace domains are thus separated by the
surface defined by $\psi=0$ which is unknown {\em a priori}; in
other words, we have to solve an elliptic partial differential
equation in a {\em free boundary} problem.

In the ordinary differential problem that characterizes the
construction of spherical models with finite mass, the condition
of vanishing cluster potential at large radii (together with the
regularity conditions at the origin) overdetermines the problem,
which can then be seen as an eigenvalue problem \citetext{e.g., see
Sect.~2.5 in \citealp{BerSti93}}. Indeed, for the King models the 
integration of the Poisson equation from the origin outwards, with 
``initial conditions" (\ref{cond1})-(\ref{cond2}), sets the relation 
between the ratio $r_{tr}/r_0$ and $\Psi$ in order to meet the 
requirement (\ref{cond3}), with $r_{tr}/r_0$ thus playing the role 
of an ``eigenvalue".

In the more complex, three-dimensional situation that we are
facing here, the existence of two different domains, internal
(Poisson) and external (Laplace), suggests the use of the method
of matched asymptotic expansions in order to obtain a uniform
solution across the separation free surface. The solution in
the internal and external domains are expressed as an asymptotic
series with respect to the tidal parameter $\epsilon$, which is
assumed to be small (following the physical model described in the
previous Section):
\begin{equation}\label{asymserin}
\psi^{(int)}({\bf \hat{r}};\epsilon)=\sum_{k=0}^{\infty}\frac{1}
{k!}\psi_k^{(int)}({\bf \hat{r}}) \epsilon^k~,
\end{equation}
\begin{equation}\label{asymserext}
\psi^{(ext)}({\bf \hat{r}};\epsilon)=\sum_{k=0}^{\infty}\frac{1}
{k!}\psi_k^{(ext)}({\bf \hat{r}})
\epsilon^k~,
\end{equation}
with spherical symmetry assumed for the zero-th order terms. The 
internal solution should obey the boundary conditions
(\ref{cond1})-(\ref{cond2}), while the external solution should
satisfy Eq.~(\ref{cond3}). The two representations should be
properly connected at the surface of the cluster.
 
On the other hand, for any small but finite value of $\epsilon$
the boundary, defined by $\psi = 0$, will be different from the
unperturbed boundary, defined by $\psi_0 = 0$, so that, for each
of the two representations given above, there will be a small
region in the vicinity of the surface of the cluster where the
leading term is vanishingly small and actually smaller than the
remaining terms of the formal series. Therefore, we expect the
validity of the expansion to break down where the second term
becomes comparable to the first, i.e where $\psi_0 =
\mathcal{O}(\epsilon)$. This region can be considered as a {\em
boundary layer}, which should be examined in ``microscopic" detail
by a suitable rescaling of the spatial coordinates and for which
an adequate solution $\psi^{(lay)}$, expressed as a different
asymptotic series, should be constructed. To obtain a uniformly
valid solution over the entire space, an asymptotic matching is
performed between the pairs ($\psi^{(int)}$, $\psi^{(lay)}$) and
($\psi^{(lay)}$, $\psi^{(ext)}$), thus leading to a solution
$\psi$, obeying all the desired boundary conditions, in terms of
three different, but matched, representations. This method of
solution is basically the same method proposed by \citet{Smi75} for
the analogous mathematical problem that arises in the
determination of the structure of rigidly rotating fluid
polytropes.

\section{Solution in terms of matched asymptotic expansions}

The complete solution to two significant orders in the tidal
parameter is now presented. The formal solution to three orders is
also displayed because of the requirements of the Van Dyke principle 
of asymptotic matching \citetext{cf. \citealp{Dyk75}, Eq.~[5.24]} that we have 
adopted.

\subsection{Internal region}

If we insert the series (\ref{asymserin}) in the Poisson equation
(\ref{Poisson}), under the conditions (\ref{cond1})-(\ref{cond2}),
we obtain an (infinite) set of Cauchy problems for $\psi_k$. The
problem for the zero-th order term (i.e., the unperturbed problem
with $\epsilon = 0$) is the one defining the construction of the
spherical and fully self-consistent King models:
\begin{equation}\label{ord0}
{\psi_0^{(int)}}^{''}+\frac{2}{\hat{r}}
{\psi_0 ^{(int)}}'=-9\frac{\hat{\rho}\left(\psi_0^{(int)}\right)}
{\hat{\rho}(\Psi)}~,
\end{equation}
with $\psi_0^{(int)}(0)=\Psi$ and ${\psi_0^{(int)}}'(0)= 0$, where 
the symbol $'$ denotes derivative with respect to the argument 
$\hat{r}$. We recall that the truncation radius $\hat{r}_{tr}$,
which defines the boundary of the spherical models, is given implicitly
by $\psi_0^{(int)}(\hat{r}_{tr})=0$.

Let us introduce the quantities:
\begin{equation}\label{rj}
R_j(\hat{r};\Psi) \equiv \frac{9}{\hat{\rho}(\Psi)}\frac{d^j
\hat{\rho}}{d\psi^j} \biggl\vert_{\psi_0^{(int)}}~.
\end{equation}
These quantities depend on $\hat{r}$ implicitly, through
the function $\psi_0^{(int)}$; in turn, the dependence on $\Psi$
is both explicit (through the term $\hat{\rho}(\Psi)$) and
implicit (because the function $\psi_0^{(int)}(\hat{r})$ depends
on the value of $\Psi$). For convenience, we give the 
expression of the first terms of the sequence (cf. Eq.~[\ref{rho}]):
$R_1 =[9/\hat{\rho}(\Psi)][\hat{\rho}(\psi_0^{(int)}) +
(\psi_0^{(int)})^{3/2}]$,
$R_2=R_1+27(\psi_0^{(int)})^{1/2}/[2\hat{\rho}(\Psi)]$,
$R_3=R_2+27(\psi_0^{(int)})^{-1/2}/[4\hat{\rho}(\Psi)]$.
Note that for $\psi_0^{(int)}\rightarrow 0$, i.e. for
$\hat{r} \rightarrow \hat{r}_{tr}$,
$R_1 \rightarrow 0$, $R_2 \rightarrow 0$, while for $j
\ge 3$ the quantities $R_j$ actually {\it diverge}. This is one
more indication of the singular character of our perturbation
analysis, which brings in some fractional power dependence on the
perturbation parameter $\epsilon$ (see also expansion [\ref{rhoexp}]).

Therefore, the equations governing the next two orders (for
$\psi_k^{(int)}$ with $k = 1,2$) can be written as:
\begin{equation}\label{ord1}
\left[\hat{\nabla}^2 +
R_1(\hat{r};\Psi)\right]\psi_1^{(int)}=-9(1-\nu)
\end{equation}
\begin{equation}\label{ord2}
\left[\hat{\nabla}^2+R_1(\hat{r};\Psi)\right]\psi_2^{(int)}=
-R_2(\hat{r};\Psi) ({\psi_1^{(int)}})^2
\end{equation}
\noindent with $\psi_1^{(int)}({\bf 0})=\psi_2^{(int)}({\bf 0})=0$
and $\hat{\nabla}\psi_1^{(int)}({\bf
0})=\hat{\nabla}\psi_2^{(int)}({\bf 0}) ={\bf 0}$. The equation
for $k=3$ is recorded in Appendix A.1, where we also describe the
structure of the general equation for $\psi_k^{(int)}$.

For any given order of the expansion, the operator acting on the
function $\psi_k^{(int)}$ (see the left-hand side of
Eqs.~[\ref{ord1}] and [\ref{ord2}]) is the same, i.e. a Laplacian
``shifted" by the function $R_1(\hat{r};\Psi)$. If we thus expand
every term $\psi_k({\bf \hat{r}})$ in spherical
harmonics\footnote{We use orthonormalized real spherical
harmonics with Condon-Shortley phase; with respect to the toroidal
angle $\phi$, they are even for $m \ge 0$ and odd otherwise.}:
\begin{equation}
\psi_k^{(int)}({\bf
\hat{r}})=\sum_{l=0}^{\infty}\sum_{m=-l}^{l}\psi_{k,lm}
^{(int)}(\hat{r})Y_{lm}(\theta,\phi)~,
\end{equation}
the three-dimensional differential problem is reduced to
a one-dimensional (radial) problem, characterized by the following
second order, linear ordinary differential operator:
\begin{equation}
\mathcal{D}_l = \frac{d^2 }{d
\hat{r}^2}+\frac{2}{\hat{r}}\frac{d}{d\hat{r}}
-\frac{l(l+1)}{\hat{r}^2}+R_1(\hat{r};\Psi)~.
\end{equation}
\noindent In general, for a fixed value of $l$, two independent
solutions to the homogeneous problem $\mathcal{D}_l f = 0$ are
expected \footnote{We note that
$R_1(0,\Psi)=9[1+\Psi^{3/2}/\hat{\rho}(\Psi)]$, i.e. a numerical positive
constant. Therefore, for $\hat{r} \rightarrow 0$ the operator $\mathcal{D}_l$
tends to the operator associated with the spherical Bessel
functions of the first and of the second kind \citetext{e.g., see
\citealp{AbrSte}, Eq.~[10.1.1] for the equation and
Eqs.~[10.1.4] and [10.1.5] for the limiting values of the functions
for small argument}.} to behave like $\hat{r}^l$ and $1/
\hat{r}^{l+1}$ for $\hat{r} \rightarrow 0$. Because of the
presence of $R_1(\hat{r};\Psi)$, solutions to equations where
$\mathcal{D}_l$ appears have to be obtained numerically.

For $k=1$ (see Eq.~[\ref{ord1}]) we thus have to address the
following problem. For $l=0$, the relevant equation is:
\begin{equation}\label{ord100}
\mathcal{D}_0 f_{00}=-9(1-\nu),
\end{equation}
where $f_{00} \equiv \psi_{1,00}^{(int)}/\sqrt{4\pi}$,
with $f_{00}(0) = f_{00}'(0) = 0$. Here we do not have to worry
about including solutions to the associated homogeneous problem,
because one of the two independent solutions would be singular at
the origin and the other would be forced to vanish by the required
condition at $\hat{r} = 0$. For $l \ge 1$ we have:
\begin{equation}\label{ord1lm}
\mathcal{D}_l\psi_{1,lm}^{(int)} = 0
\end{equation}
with $\psi_{1,lm}^{(int)}(0)={\psi_{1,lm}^{(int)}}'(0)=0$. 
Both Eq.~(\ref{ord1lm}) and the associated boundary conditions are
homogeneous. Therefore, the solution is undetermined by an
$m$-dependent multiplicative constant:
$\psi_{1,lm}^{(int)}(\hat{r})=A_{lm}\gamma_l (\hat{r})$, with
$\gamma_l(\hat{r})\sim \hat{r}^l$ for $\hat{r} \rightarrow 0$ (the
singular solution is excluded by the boundary conditions at the
origin). Then the complete formal solution is:
\begin{equation}\label{sol1}
\psi_{1}^{(int)}({\bf \hat{r}})=f_{00}(\hat{r}) +
\sum_{l=1}^{\infty}
\sum_{m=-l}^{l}A_{lm}\gamma_l(\hat{r})Y_{lm}(\theta,\phi)~,
\end{equation}
where the constants are ready to be determined by means
of the asymptotic matching with $\psi_{1}^{(lay)}({\bf \hat{r}})$
at the boundary layer.

For $k = 2$ (see Eq.~[\ref{ord2}]) the relevant equations are:
\begin{equation}\label{ord2lm}
\mathcal{D}_l\psi_{2,lm}^{(int)}=-R_2(\hat{r};\Psi)\left[{\psi_1^{(int)}}
^2\right]_{lm}~,
\end{equation}
where on the right-hand side the function $\psi_1^{(int)}$ is that 
derived from the solution of the first order problem (which shows 
the progressive character of this method for the construction of 
solutions). In Appendix A.2 the equations for the six relevant 
harmonics are displayed explicitly. The boundary conditions to be 
imposed at the origin are again homogeneous: $\psi_{2,lm}^{(int)}
(0) = {\psi_{2,lm}^{(int)}}'(0) =0$. For a fixed harmonic $(l,m)$ 
with $l > 0$, the general solution of Eq.~(\ref{ord2lm}) is the sum 
of a particular solution (which we will denote by $g_{lm} (\hat{r})$) 
and of a regular solution to the associated homogeneous problem given by
Eq.~(\ref{ord1lm}) (which we will call $B_{lm} \gamma_l(\hat{r})$,
with $ \gamma_l(\hat{r})$ the same functions introduced for the
first order problem). Obviously, the particular solution exists
only when Eq.~(\ref{ord2lm}) is non-homogeneous, i.e. only for
those values of $(l,m)$ that correspond to a non-vanishing
coefficient in the expansion of $({\psi_1^{(int)}})^2$ in spherical
harmonics. As noted in the first order problem ($k=1$), the
associated homogeneous problem for $l = 0$ has no non-trivial
solution. Then we can express the complete solution as:
\begin{equation}\label{sol2}
\psi_{2}^{(int)}({\bf \hat{r}})= g_{00}(\hat{r}) +
\sum_{l=1}^{\infty}\sum_{m=-l}^{l}[g_{lm}
(\hat{r})+B_{lm}\gamma_l(\hat{r})]Y_{lm}(\theta,\phi)~,
\end{equation}
where $B_{lm}$ are constants to be determined from the matching with the 
boundary layer.

Similarly, for $k=3$ the solution can be written as:
\begin{equation}\label{sol3}
\psi_{3}^{(int)}({\bf \hat{r}}) = h_{00}(\hat{r}) +
\sum_{l=1}^{\infty}\sum_{m=-l}^{l}
[h_{lm}(\hat{r})+C_{lm}\gamma_l(\hat{r})]Y_{lm}(\theta,\phi)~,\\
\end{equation}
where $h_{lm}$ are particular solutions and $C_{lm}$ are
constants, again to be determined from the matching with the
boundary layer.

Because the differential operator $\mathcal{D}_l$ and the boundary
conditions at the origin are the same for the reduced radial
problem of every order, we have thus obtained the general
structure of the solution for the internal region (see Appendix
A.1).

\subsection{External region}

Here we first present the general solution and then proceed to set
up the asymptotic series (\ref{asymserext}).

The solution to Eq.~(\ref{vacuum}) describing the external region,
i.e. in the Laplace domain, can be expressed as the sum of a
particular solution ($- \epsilon T({\bf \hat{r}})$) and of
the solutions to the radial part of the Laplacian operator
consistent with the boundary condition (\ref{cond3}):
\begin{equation}\label{extgen}
\psi^{(ext)}({\bf \hat{r}})=\alpha
-\frac{\lambda}{\hat{r}}-\sum_{l=1}^{\infty}\sum_{m=-l}^{l}
\frac{\beta_{lm}}{\hat{r}^{l+1}}Y_{lm}(\theta,\phi)-
\epsilon T({\bf \hat{r}})~.
\end{equation}
Here we note that the tidal potential contributes only with spherical 
harmonics of order $l=0,2$ with even values of $m$:
\begin{eqnarray}
&&T_{00}(\hat{r})=-3\sqrt \pi (\nu-1)\hat{r}^2 \label{T00}~,\\
&&T_{20}(\hat{r})=3\sqrt{ \frac{\pi}{5}} (2+\nu)\hat{r}^2 \label{T20}~,\\
&&T_{22}(\hat{r})=-3\sqrt{ \frac{3\pi}{5}} \nu \hat{r}^2 \label{T22}~.
\end{eqnarray}
At this point we can proceed to set up the asymptotic series, by
expanding the constant coefficients $\alpha$, $\lambda$, and
$\beta_{lm}$ with respect to $\epsilon$:
\begin{equation}\label{alpha}
\alpha= aH_0 =
\alpha_0+\alpha_1\epsilon+\frac{1}{2!}\alpha_2\epsilon^2+..~,
\end{equation}
\begin{equation}
\lambda=\lambda_0+\lambda_1\epsilon+\frac{1}{2!}\lambda_2\epsilon^2+..~,
\end{equation}
\begin{equation}
\beta_{lm}=a_{lm}\epsilon
+\frac{1}{2!}b_{lm}\epsilon^2+\frac{1}{3!} c_{lm}\epsilon^3+..~.
\end{equation}
The last expansion starts with a first order term because the density 
distribution of the unperturbed problem is spherically symmetric.

For convenience, we give the explicit expression of the external
solution up to third order:
\begin{eqnarray}\label{ext}
&&\psi^{(ext)}({\bf \hat{r}})=
\alpha_0-\frac{\lambda_0}{\hat{r}}+\left\{\alpha_1-\frac{\lambda_1}
{\hat{r}}-\frac{T_{00}(\hat{r})}{2\sqrt \pi}-\sum_{l=1}^{\infty}
\sum_{m=-l}^{l}\left[ \frac{a_{lm}}{\hat{r}^{l+1}}+T_{lm}(\hat{r})\right]
Y_{lm}(\theta,\phi)\right\}\epsilon\nonumber\\
&&+\frac{1}{2!}\left[\alpha_2-\frac{\lambda_2}{\hat{r}}-
\sum_{l=1}^{\infty}\sum_{m=-l}^{l}\frac{b_{lm}}{\hat{r}^{l+1}}
Y_{lm}(\theta,\phi)\right]\epsilon^2+\frac{1}{3!}\left[\alpha_3-
\frac{\lambda_3}{\hat{r}} -\sum_{l=1}^{\infty}\sum_{m=-l}^{l}
\frac{c_{lm}}{\hat{r}^{l+1}}Y_{lm}(\theta,\phi) \right]\epsilon^3~.
\end{eqnarray}

\subsection{Boundary layer}

The boundary layer is the region where the function $\psi$ becomes
vanishingly small. Since the unperturbed gravitational field at
the truncation radius is finite, $\psi_0'(\hat{r}_{tr})\ne 0$, for
any value of $\Psi$, based on a Taylor expansion of $\psi_0$ about
$\hat{r}=\hat{r}_{tr}$ we may argue that the region in which the
series (\ref{asymserin}) breaks down can be defined by
$\hat{r}_{tr}-\hat{r}= \mathcal{O}(\epsilon)$. In this boundary
layer we thus introduce a suitable change of variables:
\begin{equation}\label{eta}
\eta=\frac{\hat{r}_{tr}-\hat{r}}{\epsilon}~,
\end{equation}
take the ordering $\psi^{(lay)}=\mathcal{O}(\epsilon)$, and thus 
rescale the solution by introducing the function $\tau \equiv 
\psi^{(lay)}/\epsilon$. For positive values of $\tau$ the Poisson 
equation (\ref{Poisson}) thus becomes:
\begin{equation}\label{Poislay}
\frac{\partial^2  \tau}{\partial \eta^2}-\frac{2 \epsilon}{\hat{r}_{tr}-
\epsilon \eta}\frac{\partial \tau}{\partial \eta}+\frac{\epsilon^2}
{(\hat{r}_{tr}-\epsilon \eta)^2}\Lambda^2\tau=
-\frac{9}{\hat{\rho}(\Psi)}\,\epsilon\hat{\rho}
(\epsilon \tau)-9\epsilon^2(1-\nu)~,
\end{equation}
where $\Lambda^2$ is the angular part of the Laplacian in spherical 
coordinates. For negative values of $\tau$ we can write a similar 
equation, corresponding to Eq.~(\ref{vacuum}), which is obtained 
from Eq.~(\ref{Poislay}) by dropping the term proportional to 
$\hat{\rho}(\epsilon \tau)$.

With the help of the asymptotic expansion for small argument of
the incomplete gamma function \citetext{e.g., see \citealp{BenOrs99},
Eq.~[6.2.5]}, we find:
\begin{equation}\label{rhoexp}
\hat{\rho}(\epsilon\tau) \sim \frac{2}{5}\tau^{5/2}\epsilon^{5/2}
+\frac{4}{35}\tau^{7/2}\epsilon^{7/2}+...~,
\end{equation}
so that, within the boundary layer, the contribution of
$\hat{\rho}(\epsilon \tau)$ (which is the one that distinguishes
the Poisson from the Laplace regime) becomes significant only
beyond the tidal term, as a correction
$\mathcal{O}(\epsilon^{7/2})$.

Therefore, up to $\mathcal{O}(\epsilon^{2})$ we can write:
\begin{equation}\label{tau}
\tau= \tau_0 + \tau_1 \epsilon + \frac{1}{2!}\tau_2\epsilon^2.
\end{equation}
To this order, which is required for a full solution up
to $k=2$ of the global problem (see Eqs.~[\ref{asymserin}] and 
[\ref{asymserext}]), by equating in Eq.~(\ref{Poislay}) the first 
powers of $\epsilon$ separately, we obtain the relevant equations
for the first three terms:
\begin{equation}
\frac{\partial^2 \tau_0}{\partial \eta^2}=0~,
\end{equation}
\begin{equation}
\frac{\partial^2 \tau_1}{\partial \eta^2}=\frac{2}{\hat{r}_{tr}}\frac
{\partial \tau_0}{\partial \eta}~,
\end{equation}
\begin{equation}
\frac{\partial^2 \tau_2}{\partial
\eta^2}=\frac{4}{\hat{r}_{tr}}\left[ \frac{\partial
\tau_1}{\partial \eta}+\frac{\eta}{\hat{r}_{tr}}\frac {\partial
\tau_0}{\partial \eta}\right]-\frac{2}{\hat{r}_{tr}^2}\Lambda^2
\tau_0-18(1-\nu)~.
\end{equation}
The equations are easily integrated in the variable $\eta$, to 
obtain the solutions:
\begin{equation}\label{tau0}
\tau_0=F_0(\theta,\phi)\eta+G_0(\theta,\phi)~,
\end{equation}
\begin{equation}\label{tau1}
\tau_1=\frac{F_0(\theta,\phi)}{\hat{r}_{tr}}\eta^2+F_1(\theta,\phi)\eta
+G_1(\theta,\phi)~,
\end{equation}
\begin{eqnarray}\label{tau2}
&&\tau_2=\frac{2 F_0(\theta,\phi)}{\hat{r}_{tr}^2}\eta^3-\frac{1}
{3 \hat{r}_{tr}^2}\Lambda^2 F_0(\theta,\phi)\eta^3+\frac{2F_1
(\theta,\phi)}{\hat{r}_{tr}}\eta^2\nonumber\\
&&-9(1-\nu)\eta^2-\frac{1}{\hat{r}_{tr}^2}\Lambda^2
G_0(\theta,\phi)\eta^2+F_2(\theta,\phi)\eta+G_2(\theta,\phi)~.
\end{eqnarray}
The six free angular functions that appear in the formal solutions 
will be determined by the matching procedure.

\subsection{Asymptotic matching to two orders}

In order to obtain the solution, we must perform separately the 
relevant matching for the pairs $(\psi^{(int)},\psi^{(lay)})$ and 
$(\psi^{(lay)},\psi^{(ext)})$. We follow the Van Dyke matching
principle, which requires that we compare the second order expansion 
of the internal and external solutions  with the third order expansion 
of the boundary layer solution. The full procedure is described in 
Appendix A.3.

To first order (i.e., up to $k=1$ in series [\ref{asymserin}] and
[\ref{asymserext}]), from the matching of the pair $(\psi^{(int)},
\psi^{(lay)})$ we find the free angular functions of (\ref{tau0}) 
and (\ref{tau1}):
\begin{equation}\label{F0}
 F_0(\theta,\phi)=-{\psi_0^{(int)}}'(\hat{r}_{tr})~,
\end{equation}
\begin{equation}\label{G0}
G_0(\theta,\phi)=\psi_1^{(int)}(\hat{r}_{tr},\theta,\phi)~,
\end{equation}
\begin{equation}\label{F1}
F_1(\theta,\phi)=-\frac{\partial \psi_1^{(int)}}{\partial \hat{r}}
(\hat{r}_{tr},\theta,\phi)~,
\end{equation}
\begin{equation}\label{G1}
G_1(\theta,\phi)=\frac{1}{2}\psi_2^{(int)}(\hat{r}_{tr},\theta,\phi)~.
\end{equation}
From the matching of the pair $(\psi^{(ext)},\psi^{(lay)})$ we 
connect $\psi^{(ext)}$ to the same angular functions, thus proving 
that the matching to first order is equivalent to imposing continuity 
of the solution up to second order and of the first radial derivative 
up to first order. This allows us to determine the free constants that 
are present in the first two terms of (\ref{ext}) and in (\ref{sol1}):
\begin{equation}\label{alpha0}
\alpha_0=\frac{\lambda_0}{\hat{r}_{tr}}~,
\end{equation}
\begin{equation}\label{lambda0}
\lambda_0=\hat{r}_{tr}^2{\psi_0^{(int)}}'(\hat{r}_{tr})~,
\end{equation}
\begin{equation}\label{alpha1}
\alpha_1=f_{00}(\hat{r}_{tr})+\hat{r}_{tr}f_{00}'(\hat{r}_{tr})+\frac
{3 T_{00}(\hat{r}_{tr})}{2\sqrt \pi}~,
\end{equation}
\begin{equation}\label{lambda1}
\lambda_1=\hat{r}_{tr}^2f_{00}'(\hat{r}_{tr})+\frac{\hat{r}_{tr}T_{00}
(\hat{r}_{tr})}{ \sqrt\pi}~,
\end{equation}
\begin{equation}\label{A2m}
A_{2m}=-\frac{5 T_{2m}(\hat{r}_{tr})}{\hat{r}_{tr}\gamma_2'(\hat{r}
_{tr})+3\gamma_2(\hat{r}_{tr})}~,
\end{equation}
\begin{equation}\label{a2m}
a_{2m}=-\hat{r}_{tr}^3[A_{2m}\gamma_2(\hat{r}_{tr})+
T_{2m}(\hat{r}_{tr})]~.
\end{equation}
Note that $A_{lm}=a_{lm}=0$ if $l\ne 2$, for every value of $m$, and that 
the constants for $l=2$ are non-vanishing only for $m=0,2$. The constants 
that identify the solution are thus expressed in terms of the values of the 
unperturbed field ${\psi_0^{(int)}}'$, of the ``driving" tidal potential 
$T_{lm}$, and of the solutions $f_{00}$ and $\gamma_2$ (see Eqs.~[\ref{ord100}] 
and [\ref{ord1lm}]) taken at $\hat{r} = \hat{r}_{tr}$.

The boundary surface of the first order model is defined implicitly by 
$\psi_0^{(ext)}(\hat{r})+\psi_1^{(ext)}(\hat{r},\theta,\phi)\epsilon=0$,
i.e. the spherical shape of the King model is modified by monopole and 
quadrupole contributions, which are even with respect to toroidal and 
poloidal angles and characterized by reflection symmetry with respect to 
the three natural coordinates planes. As might have been expected from 
the physical model, the spherical shape is thus modified only by spherical 
harmonics $(l,m)$ for which the tidal potential has non-vanishing coefficients. 
Mathematically, this is non-trivial, because the first order equation
in the internal region Eq.~(\ref{ord1}) is non-homogeneous only for $l=0$; 
the quadrupole contribution to the internal solution is formally ``hidden" 
by the use of the function $\psi$ (which includes the tidal potential) and
is unveiled by the matching which demonstrates that $A_{2m}$
with $m=0,2$ are non-vanishing.

The first order solution can be inserted into the right-hand side of
Eq.~(\ref{ord2lm}) to generate non-homogeneous equations (and thus 
particular solutions) only for $l=0,2,4$ and corresponding positive 
and even values of $m$ (see Appendix A.2). We can thus proceed to 
construct the second order solution in the same way described above
for the first order solution. From the matching of the pair 
$(\psi^{(int)},\psi^{(lay)})$ we determine the missing angular 
functions:
\begin{equation}\label{F2}
F_2(\theta,\phi)=-\frac{\partial \psi_2^{(int)}}{\partial \hat{r}}
(\hat{r}_{tr},\theta,\phi)~,
\end{equation}
\begin{equation}\label{G2}
G_2(\theta,\phi)=\frac{1}{3}\psi_3^{(int)}(\hat{r}_{tr},\theta,\phi)~,
\end{equation}
which are then connected to the properties of $\psi^{(ext)}$ by the 
matching of the pair $(\psi^{(ext)},\psi^{(lay)})$. This is equivalent 
to imposing continuity of the solution up to third order and of the 
first radial derivative up to second order and leads to the determination 
of the free constants that appear in the third term of (\ref{ext}) and 
in (\ref{sol2}):
\begin{equation}\label{alpha2}
\alpha_2=g_{00}(\hat{r}_{tr})+\hat{r}_{tr}g_{00}'(\hat{r}_{tr})~,
\end{equation}
\begin{equation}\label{lambda2}
\lambda_2=\hat{r}_{tr}^2g_{00}'(\hat{r}_{tr})~,
\end{equation}
\begin{equation}\label{B2m}
B_{2m}=-\frac{\hat{r}_{tr} g_{2m}'(\hat{r}_{tr})+3g_{2m}(\hat{r}_{tr})}
{\hat{r}_{tr}\gamma_2'(\hat{r}_{tr})+3\gamma_2(\hat{r}_{tr})}~,
\end{equation}
\begin{equation}\label{b2m}
b_{2m}=-\hat{r}_{tr}^3[g_{2m}(\hat{r}_{tr} )+B_{2m} \gamma_2(\hat{r}_{tr})]~,
\end{equation}
\begin{equation}\label{B4m}
B_{4m}=-\frac{\hat{r}_{tr} g_{4m}'(\hat{r}_{tr})+5g_{4m}(\hat{r}_{tr})}
{\hat{r}_{tr}\gamma_4'(\hat{r}_{tr})+5\gamma_4(\hat{r}_{tr})}~,
\end{equation}
\begin{equation}\label{b4m}
b_{4m}=-\hat{r}_{tr}^5[g_{4m}(\hat{r}_{tr} )+B_{4m} \gamma_4(\hat{r}_{tr})]~.
\end{equation}
Here $B_{lm}=b_{lm}=0$ if $l\ne 2,4$ for every value of $m$; the only 
non-vanishing constants with $l=2,4$ are those with even $m$.

Therefore, the second order solution has non-vanishing contributions only
for $l=0, 2,4$, i.e. for those harmonics for which the particular solution to 
Eq.~(\ref{ord2}) is non-trivial. {\it By induction}, it can be proved (see 
Appendix A.4) that the $k$-th order solution is characterized by $l=0,2,..,2k$
harmonics with corrisponding positive and even values of $m$.  In reality, 
the discussion of the matching to higher orders ($k>3$) would require a 
re-definition of the boundary layer, because the density contribution on the 
right-hand side of Eq.~(\ref{Poislay}) (for positive values of $\tau$) comes 
into play.  The asymptotic matching procedure carries through also in this 
more complex case but, for simplicity, is omitted here. We should also keep 
in mind that in an asymptotic analysis the inclusion of higher order terms 
does not necessarily lead to better accuracy in the solution; the optimal 
truncation in the asymptotic series depends on the value of the expansion 
parameter (in this case, on the value of $\epsilon$) and has to be judged 
empirically.

In conclusion, starting from a given value of the King concentration parameter 
$\Psi$ and from a given strength of the tidal field $\epsilon$, the uniform 
triaxial solution is constructed by numerically integrating Eqs.~(\ref{ord0}),
(\ref{ord100}), (\ref{ord1lm}), and (\ref{ord2lm}) and by applying the 
constants derived in this subsection to the asymptotic series expansion 
(\ref{asymserin})-(\ref{asymserext}). The numerical integrations can be 
performed efficiently by means of standard Runge-Kutta routines. The boundary 
surface of the model is thus defined by $\psi_0^{(ext)}(\hat{r})+\psi_1^{(ext)}
(\hat{r},\theta,\phi)\epsilon + \psi_2^{(ext)}(\hat{r},\theta,\phi)\epsilon^2/2=0$, 
while the internal density distribution is given by $\rho = \rho(\psi_0^{(int)}
(\hat{r})+\psi_1^{(int)}(\hat{r},\theta,\phi)\epsilon + \psi_2^{(int)}(\hat{r},
\theta,\phi)\epsilon^2/2)$, with the function $\rho$ defined by Eq.~(\ref{rho}). 
Any other ``observable" quantity can be reconstructed by suitable integration in 
phase space of the distribution function $f_K(H)$ defined by Eq.~(\ref{fK}), with 
$H$ defined by Eq.~(\ref{Jacobi}), and $\Phi_T + \Phi_C = H_0 - [\psi_0^{(int)}
(\hat{r})+\psi_1^{(int)}(\hat{r},\theta,\phi)\epsilon + \psi_2^{(int)}(\hat{r},
\theta,\phi)\epsilon^2/2]/a$.

In Fig.~1 and Fig.~2 we illustrate the main characteristics of one 
triaxial model constructed with the method described in this Section.

\section{Alternative methods of solution}
 
\subsection{The method of strained coordinates}
 
The mathematical problem described in Sect.~3 can also be solved
by the method of {\em strained coordinates}, an alternative method
usually applied to
non-linear hyperbolic differential equations \citetext{e.g., see 
\citealp{Dyk75}, Chapter 6} and considered by \citet{Smi76} in the 
solution of the singular free-boundary perturbation problem that 
arises in the study of rotating polytropes.
 
Starting from a series representation of the form
(\ref{asymserin}) and (\ref{asymserext}) for the solution defined in
the Poisson and Laplace domains, respectively, a transformation is
considered from spherical coordinates $(\hat{r},\theta,\phi)$ to
``strained coordinates" $(s,p,q)$:
\begin{eqnarray}\label{trans}
&& \hat{r}=s+\epsilon \hat{r}_1(s,p,q)+\frac{1}{2} \epsilon^2
\hat{r}_2(s,p,q)+... \nonumber \\ &&\theta=p \nonumber \\
&&\phi=q~,
\end{eqnarray}
where $\hat{r}_k(s,p,q)$ are initially unspecified straining functions. 
We note that the zero-th order problem is defined by the same 
Eq.~(\ref{ord0}) with the same boundary conditions but with the variable 
$\hat{r}$ replaced by $s$. The unperturbed spherical boundary in the 
strained space is defined by $s=s_0$, where $\psi_0^{(int)}(s_0)=0$. 
To each order, the effective boundary of the perturbed configuration 
remains described by the surface $s=s_0$, while in physical coordinates
the truncation radius actually changes as a result of the straining 
functions $\hat{r}_k$ that are determined progressively.
 
The Laplacian expressed in the new coordinates, $\widetilde{\nabla}^2$, 
can be written as an asymptotic series: $\widetilde{\nabla}^2= L_0+
\epsilon L_1+1/2 \epsilon^2 L_2+...$, where $L_k$ are linear second order 
operators \footnote{Surfaces with constant $s$ in the strained space are 
assumed to correspond to surfaces with constant $\psi^{(int)}$ in the 
physical space, i.e. $\psi^{(int)}=\psi^{(int)}(s)$; therefore, $L_k$ 
(with $k \ge 0$) is an ordinary differential operator for $\psi^{(int)}$.} 
in which $\hat{r}_j(s,p,q)$ (with $j=1,..,k$) and their derivatives appear. 
For convenience, we record the zero-th and first order operators:
\begin{equation}
L_0\equiv \frac{d ^2}{d s^2}+ \frac{2}{s}\frac{d}{d s}~,
\end{equation}
\begin{equation}\label{L1}
 L_1\equiv -\left(2\frac{\partial \hat{r}_1}{\partial s}\right)
\frac{d^2}{d s^2}-\left(\frac{\partial^2 \hat{r}_1}{\partial s^2}+
\frac{2}{s}\frac{\partial\hat{r}_1}{\partial s} + \frac{1}{s^2}\Lambda^2
\hat{r}_1+\frac{2}{s^2}\hat{r}_1\right) \frac{d }{d s}~,
\end{equation}
where $\Lambda^2$ is the standard angular part of the Laplacian, written 
with angular coordinates $(p,q)$.  The general $k$-th order operator can 
be decomposed as $L_k=L_1+F_k$, where $F_k$ is, in turn, a second order 
operator in which $\hat{r}_j(s,p,q)$ (with $j=1,..,k-1$) appear and $L_1$ 
is defined as in Eq.~(\ref{L1}) but with $\hat{r}_k(s,p,q)$ instead of 
$\hat{r}_1(s,p,q)$; these operators appear in the relevant equation for 
$\psi_k^{(int)}$:
\begin{equation}
[L_0+R_1(\psi;\epsilon)] \psi_k^{(int)}=L_k\psi_0^{(int)}
\end{equation}  
which corresponds to the general $k$-th order equation of the previous method.
 
Following a set of constraints that guarantee the regularity of the 
series (\ref{asymserin}) in the strained space, the equations that 
uniquely identify the straining functions to any desired order can 
be found and solved numerically; structurally, they somewhat correspond 
to Eqs.~(\ref{ord1lm}) and (\ref{ord2lm}) in Sect.~4.1. Therefore, the 
internal and external solutions can be worked out and patched by requiring 
continuity of the solution and of the first derivative with respect to 
the variable $s$ at the boundary surface defined by $s=s_0$, in general 
qualitative analogy with the method described in the previous section.
 
This method is formally more elegant than the method of matched
asymptotic expansions but requires a more significant
numerical effort because, even though the number of equations to
be solved at each order is the same, the operator that plays here
a central role in the equations for the straining functions, 
$L_1\psi_0^{(int)}$ (interpreted as an operator acting on $\hat{r}_k[s,p,q]$), 
is more complex than $\mathcal{D}_l$ (defined in Sect.~4.1).
 
\subsection{Iteration}
 
This technique follows the approach taken by \citet{PreTom70} 
and by \citet{Wil75}, for the construction of
self-consistent dynamical models of differentially rotating
elliptical galaxies, and later by \citet{LonLag96},
for their extension of King models to the rotating case.
 
In terms of the function:
\begin{equation}
u({\bf \hat{r}}) \equiv a[H_0-\Phi_C({\bf \hat{r}})]= \psi({\bf
\hat{r}})+ \epsilon T({\bf \hat{r}})~,
\end{equation}
inside the cluster the Poisson equation can be written as:
\begin{equation}\label{iterPoi}
\hat{\nabla}^2 u=-\frac{9}{\hat{\rho}(\Psi)}\hat{\rho}(u-\epsilon T)~,
\end{equation}
while outside the cluster the Laplace equation is simply:
\begin{equation}\label{iterLap}
\hat{\nabla}^2 u=0~.
\end{equation}
The boundary conditions at the origin are $u({\bf0})=\Psi$ and 
$\hat{\nabla}u({\bf 0})={\bf 0}$, because the tidal potential 
$T({\bf \hat{r}})$ is a homogeneous function; the condition at 
large radii is $u \rightarrow aH_0$.
 
The basic idea is to get an improved solution $u^{(n)}$ of the
Poisson equation by evaluating the ``source term" on the right-hand
side with the solution obtained in the immediately previous step:
\begin{equation}\label{iterPoin}
\hat{\nabla}^2
u^{(n)}=-\frac{9}{\hat{\rho}(\Psi)}\hat{\rho}(u^{(n-1)}-\epsilon T)~.
\end{equation}
The iteration is seeded by inserting as $u^{(0)}$, on the right-hand
side of Eq.~(\ref{iterPoin}), the spherical solution of the King models. 
The iteration continues until convergence is reached.
 
In order to solve Eq.~(\ref{iterPoin}), we expand in spherical
harmonics the solution and, correspondingly, the dimensionless
density distribution:
\begin{equation}\label{uexp}
u^{(n)}({\bf \hat{r}
})=\sum_{l=0}^{\infty}\sum_{m=-l}^{l}u_{lm}^{(n)}
(\hat{r})Y_{lm}(\theta,\phi)~,
\end{equation}
\begin{equation}\label{rhoitexp}
\hat{\rho}^{(n)}({\bf \hat{r}
})=\sum_{l=0}^{\infty}\sum_{m=-l}^{l}\hat
{\rho}_{lm}^{(n)}(\hat{r})Y_{lm}(\theta,\phi)~,
\end{equation}
so that the reduced radial problems for the functions $u_{lm}^{(n)}
(\hat{r})$ are:
\begin{equation}\label{equlm}
\left[\frac{d^2 }{ d \hat{r}^2}+\frac{2}{\hat{r}}{d \over d
\hat{r}} -{l(l+1)\over \hat{r}^2}
\right]u_{lm}^{(n)}=-\frac{9}{\hat{\rho}(\Psi)}
\hat{\rho}_{lm}^{(n-1)}~,
\end{equation}
with boundary conditions $u_{00}^{(n)}(0)=\Psi$, $u_{lm}^{(n)}(0)=0$ 
and ${u_{00}^{(n)}}'(0)={u_{lm}^{(n)}}'(0)=0$.
Here, in contrast with the structure of the governing equations
for $\psi_k^{(int)}$ of Sect.~4.1 and Sect.~4.2, the radial part of 
the Laplacian appears with no ``shift", for which the homogeneous
solutions are known analytically. Thus the full solution to
Eq.~(\ref{equlm}) can be obtained in integral form by the standard
method of {\em variation of the arbitrary constants}:
\begin{equation}\label{u00}
u_{00}^{(n)}(\hat{r})=\Psi -\frac{9}{\hat{\rho}(\Psi)}\left[
\int_0^{\hat{r}}\hat{r}'\hat{\rho}_{00}^{(n-1)}
(\hat{r}')d\hat{r}'-\frac{1}{\hat{r}}\int_0^{\hat{r}}
\hat{r}'^2\hat{\rho}_{00}^{(n-1)}(\hat{r}')d\hat{r}'\right]~,
\end{equation}
\begin{equation}\label{ulm}
u_{lm}^{(n)}(\hat{r})= \frac{9}{(2l+1)\hat{\rho}(\Psi)}\left[\hat{r}^{l}
\int_{\hat{r}}^{\infty}
\hat{r}'^{1-l}\hat{\rho}_{lm}^{(n-1)}(\hat{r}')d\hat{r}'
+\frac{1}{\hat{r}^{l+1}}\int_0^{\hat{r}}
\hat{r}'^{l+2} \hat{\rho}_{lm}^{(n-1)}(\hat{r}')d\hat{r}'\right]~.
\end{equation}
The complete calculation can be found in the Appendix of \citet{PreTom70}. 
Here we only remark that this integral form is valid in both Poisson and 
Laplace domains because it contains simultaneously the regular and the 
singular homogeneous solutions of the Laplacian. In the derivation, all 
the boundary conditions have been used; in particular, the two conditions 
at the origin are sufficient to obtain expression (\ref{u00}), while for 
expression (\ref{ulm}) the one concerning the radial derivative at the origin 
is used together with the one that describes the behavior at large radii (i.e. 
$u_{lm}^{(n)}\rightarrow 0$ for $l \ge 1$). Furthermore, from the condition 
at large radii evaluated for the harmonic $l=0$, i.e. $u_{00}^{(n)}/\sqrt{4\pi}
\rightarrow aH_0^{(n)}$ (here the notation reminds us that the value of $H_0$ 
is known only approximately and it changes slightly at every iteration), we find:
\begin{equation}
aH_0^{(n)}\equiv \frac{\Psi}{\sqrt{4\pi}} -\frac{9}{\sqrt{4\pi} 
\hat{\rho}(\Psi)}\int_0^{\infty}\hat{r}'
\hat{\rho}_{00}^{(n-1)}(\hat{r}')d\hat{r}'~,
\end{equation}
where we should recall that beyond a certain radius $\hat{\rho}_{00}^{(n-1)}$ 
vanishes. 

In terms of the function $u$, the boundary of the cluster is given implicitly 
by: $u({\bf \hat{r}})=\epsilon T({\bf \hat{r}})$. Therefore, the radial 
location at which the $\hat{\rho}_{lm}^{(n-1)}$ vanishes is determined 
numerically from:
\begin{equation}
 \hat{\rho}_{lm}^{(n-1)}(\hat{r})=\int_0^{2 \pi}\int_{-1}^{1} \hat{\rho}
[u^{(n-1)}(\hat{r},\theta,\phi)-\epsilon T(\hat{r},\theta,\phi)]Y_{lm}
(\theta, \phi) d(cos \theta)d\phi~.
\end{equation}
 
In practice, to perform the iteration, the definition of a
grid in spherical coordinates and of a suitable algorithm, in order
to perform the expansion and the resummation in spherical harmonics 
of $u$ and $\hat{\rho}$, is required; the number of angular points 
of the grid and the maximum harmonic indices $(l,m)$ admitted in 
the series (\ref{uexp}) and (\ref{rhoitexp}) are obviously related.
 
\section{Conclusions}
 
Spherical King models are physically justified models of
quasi-relaxed stellar systems with a truncation radius argued to
``summarize" the action of an external tidal field. Such simple
models have had great success in representing the structure and
dynamics of globular clusters, even though the presence of the
tidal field is actually ignored. Motivated by these considerations
and by the recent major progress in the observations of globular
clusters, in this paper we have developed a systematic procedure
to construct self-consistent non-spherical models of quasi-relaxed
stellar systems, with special attention to models for which the
non-spherical shape is due to the presence of external tides.
 
The procedure developed in this paper starts from a distribution
function identified by replacing, in a reference spherical model,
the single star energy with the relevant Jacobi integral, thus
guaranteeing that the collisionless Boltzmann equation is
satisfied. Then the models are constructed by solving the Poisson
equation, an elliptic partial differential equation with free
boundary. The procedure is very general and can lead to the
construction of several families of non-spherical
equilibrium models. In particular, we have obtained the following
results:
 
\begin{itemize}
\item We have constructed models of quasi-relaxed
triaxial stellar systems in which the shape is due to the presence
of external tides; these models reduce to the standard spherical
King models when the tidal field is absent.
\item For these models we have outlined the general properties of
the relevant parameter space; in a separate paper (Varri \&
Bertin, in preparation) we will provide a thorough description of
this two-parameter family of models, also in terms of projected
quantities, as appropriate for comparisons with the observations.
\item We have given a full, explicit solution to two orders in the 
tidal strength parameter, based on the method of matched asymptotic 
expansions; by comparison with studies of analogous problems in the 
theory of rotating polytropic stars, this method appears to be most
satisfactory.
\item We have also discussed two alternative methods of
solution, one of which is based on iteration seeded by the
spherical solution; together with the use of dedicated $N$-body
simulations, the ability to solve such a complex mathematical
problem in different ways will allow us to test the quality of the
solutions in great detail.
\item By suitable change of notation and physical re-interpretation,
the procedure developed in this paper can be applied to the
construction of non-spherical quasi-relaxed stellar systems
flattened by rotation (see Appendix B).
\item The same procedure can also be applied to extend to the triaxial 
case other isotropic truncated models (such as low-$n$ polytropes), 
that is models that do not reduce to King models in the absence of
external tides (see Appendix C).
\end{itemize}
 
We hope that this contribution, in addition to extending the class
of self-consistent models of interest in stellar dynamics, will be
the basis for the development of simple quantitative tools to
investigate whether the observed shape of globular clusters is
primarily determined by internal rotation, by external tides, or
by pressure anisotropy.
 
\acknowledgments

\appendix

\section{Details of the solution in terms of matched asymptotic expansion}
 
\subsection{The general equation}
 
From the Taylor expansion about $\epsilon=0$ of the right-hand
side of Eq. (\ref{Poisson}), the structure of the equations for
$\psi_k^{(int)}$ (with $k\ge 2$) can be expressed as:
\begin{equation}\label{gen}
\left[\hat{\nabla}^2+R_1(\hat{r};\Psi)\right]\psi_k^{(int)}=-\left(
\sum_{j=2}^{k}R_j(\hat{r};\Psi)\,X_{k,j}\right)~,
\end{equation}
where $X_{k,j}$ denotes the terms that arise from the
derivatives of $\psi^{(int)}(\hat{r};\epsilon)$ with respect to
$\epsilon$, thus expressed as products of $\psi_i^{(int)}$ (with
$i=1 ,..,k-1$). For fixed $k$ and $j$, the quantity $X_{k,j}$ is
thus a sum of products of $\psi_i^{(int)}$ with subscripts that
are j-part partitions of the integer $k$. Each product of
$\psi_i^{(int)}$ is multiplied by a numerical factor defined as
the ratio between $k!$ and the factorials of the integers that are
parts of the associated partition (if an integer appears $m$ times
in the partition, the factor must also be divided by $m!$). In
particular, for $k=3$ we have:
\begin{equation}
X_{3,2}=3\psi_2^{(int)}\psi_1^{(int)}~{\rm and} \;\;\;\;
X_{3,3}=(\psi_1^{(int)})^3~,
\end{equation}
because the 2-part partition of 3 is $2+1$ and the
3-part partition is trivially $1+1+1$, thus the relevant equation
is:
\begin{equation}
\left[\hat{\nabla}^2+R_1(\hat{r};\Psi)\right]\psi_3^{(int)}=
-R_2(\hat{r};\Psi)\,3\psi_2^{(int)}\psi_1^{(int)}-R_3(\hat{r};\Psi)\,
(\psi_1^{(int)})^3~.
\end{equation}
Therefore, this formulation of the right-hand side of
the general equation (together with the term $R_1(\hat{r};\Psi)
\psi_k^{(int)}$ on the left-hand side) brings in the {\em Fa\'{a} 
di Bruno formula} \citep{Faa1855} for the $k$-th order derivative
of a composite function in which the inner one is expressed as a
series in the variable with respect to which the derivation is
performed.
 
\subsection{The equation for the second order radial problem}
 
The expansion in spherical harmonics of $(\psi_1^{(in)})^2$, which
involves the product of two spherical harmonics with $l=0$ or
$l=2$ (with $m$ positive and even), can be performed by means of
the so-called 3-j Wigner symbols \footnote{For the definition of 
3-j Wigner symbols and the expression of the harmonic expansion 
of the product of two spherical harmonics, see, e.g., \citet{Edm60}, 
Eqs.~(3.7.3) and (4.6.5), respectively.}. Equation (\ref{ord2lm}) 
thus corresponds to the following set of six equations:
\begin{eqnarray}
&&\mathcal{D}_0\psi_{2,00}^{(int)}=-R_2(\hat{r};\Psi){1\over
2\sqrt{\pi}}\;[(\psi_{1,00}^{(int)})^2+(\psi_{1,20}^{(int)})^2+
(\psi_{1,22}^{(int)})^2]~,\\
&&\mathcal{D}_2\psi_{2,20}^{(int)}=-R_2(\hat{r};\Psi){1 \over
7}\sqrt{{5\over\pi}}\;\left[{7\over
\sqrt{5}}\psi_{1,00}^{(int)}\,\psi_{1,20}^{(int)}+(\psi_{1,20}
^{(int)})^2-(\psi_{1,22}^{(int)})^2
\right]~,\\
&&\mathcal{D}_2\psi_{2,22}^{(int)}=-R_2(\hat{r};\Psi){1\over
\sqrt{\pi}}\;\left[\psi_{1,00}^{(int)}\,\psi_{1,22}^{(int)}-
{2\sqrt{5}\over 7}\psi_{1,20}^{(int)}\,\psi_{1,22}^{(int)}\right]~,\\
&&\mathcal{D}_4\psi_{2,40}^{(int)}=-R_2(\hat{r};\Psi){1\over
7\sqrt{\pi}}\;\left[3(\psi_{1,20}^{(int)})^2+\frac{1}{2}(\psi_{1,22}
^{(int)})^2\right]~,\\
&&\mathcal{D}_4\psi_{2,42}^{(int)}=-R_2(\hat{r};\Psi){1\over
7}\sqrt{{15\over\pi}}\;\psi_{1,20}^{(int)}\,\psi_{1,22}^{(int)}~,\\
&&\mathcal{D}_4\psi_{2,44}^{(int)}=-R_2(\hat{r};\Psi)\frac{1}{2}
\sqrt{{5\over 7\,\pi}}\;(\psi_{1,22}^{(int)})^2~.
\end{eqnarray}

\subsection{The asymptotic matching for the first order solution}

To derive the first order solution, the matching between the pairs
$(\psi^{(int)}, \psi^{(lay)})$ and $(\psi^{(ext)},\psi^{(lay)})$
requires that the internal (external) solution is expanded in a
Taylor series around $\hat{r} = \hat{r}_{tr}$ up to terms of order
$\mathcal{O}((\hat{r} - \hat{r}_{tr})^2)$, expressed with scaled
variables, expanded up to $\mathcal{O}(\epsilon^2)$, and
re-expressed with non-scaled variables:
\begin{eqnarray}\label{intextexp}
&&[\psi^{(~)}]^{(2)}(\hat{r},\theta,\phi)=\psi_0^{(~)}
(\hat{r}_{tr})-{\psi_0^{(~)}}'(\hat{r}_{tr})(\hat{r}_{tr}-\hat{r})
+\frac{1}{2}{\psi_0^{(~)}}^{''}(\hat{r}_{tr})(\hat{r}_{tr}-\hat{r})^2
\nonumber\\ &&+\left[\psi_1^{(~)}(\hat{r}_{tr},\theta,\phi)
-\frac{\partial \psi_1^{(~)}} {\partial
\hat{r}}(\hat{r}_{tr},\theta,\phi)(\hat{r}_{tr}-\hat{r})\right]
\epsilon+\frac{1}{2}\psi_2^{(~)}(\hat{r}_{tr},\theta,\phi)\epsilon^2~.
\end{eqnarray}
Here the closed parentheses include either ``int" or``ext", to denote 
internal or external solution, while the notation $[\;]^{(2)}$ on the 
left-hand side indicates that a second order expansion in $\epsilon$ 
has been performed.

The boundary layer solution in the vicinity of $\eta = 0$ up to
$\mathcal{O}(\eta^2)$, expressed with non-scaled variables and
expanded (formally) up to third order in $\epsilon$ is given by:
\begin{eqnarray}\label{layexp}
&& [\psi^{(lay)}]^{(3)}(\hat{r},\theta,\phi)=F_0(\theta,\phi)(\hat{r}_{tr}-\hat{r})
+\frac{F_0(\theta,\phi)}{\hat{r}_{tr}}
(\hat{r}_{tr}-\hat{r})^2 \nonumber\\
&&+[G_0(\theta,\phi)+F_1(\theta,\phi)
(\hat{r}_{tr}-\hat{r})]\epsilon+G_1(\theta,\phi)\epsilon^2~.
\end{eqnarray}
\noindent By equating equal powers of $\epsilon$ and
$(\hat{r}_{tr}-\hat{r})$ in (\ref{intextexp}) and (\ref{layexp}),
we find:
\begin{eqnarray}
 \psi^{(int)}_0(\hat{r}_{tr})&=\;\;\;\;\;0\;\;\;\;\;=&\alpha_0-
\frac{\lambda_0}{\hat{r}_{tr}}~, \label{match0}\\
{-\psi^{(int)}_0}^{'}(\hat{r}_{tr})&=F_0(\theta,\phi)=&-
\frac{\lambda_0}{\hat{r}_{tr}^2}~, \label{match1}\\
\frac{1}{2}{\psi^{(int)}_0}^{''}(\hat{r}_{tr})&\displaystyle =
\frac{F_0(\theta,\phi)}{\hat{r}_{tr}}=&-\frac{\lambda_0}
{\hat{r}_{tr}^3}~, \label{match2}\\
 \psi^{(int)}_1(\hat{r}_{tr},\theta,\phi)&=G_0(\theta,\phi)=&
\alpha_1-\frac{\lambda_1}{\hat{r}_{tr}}-\frac{T_{00}(\hat{r}_{tr})}
{2\sqrt \pi}\nonumber\\ &&-\sum_{l=1}^{\infty}\sum_{m=-l}^{l}\left[\frac{a_{lm}}
{\hat{r}^{l+1}_{tr}}+T_{lm}(\hat{r}_{tr})\right]Y_{lm}(\theta,\phi)~, \label{match3}\\
-\frac{\partial \psi_1^{(int)}}{\partial\hat{r}}(\hat{r}_{tr},\theta,\phi)&=
F_1(\theta,\phi)=&-\frac{\lambda_1}{\hat{r}^2_{tr}}+\frac{T_{00}(\hat{r}_{tr})}
{\sqrt \pi\hat{r}_{tr}}\nonumber\\ &&-\sum_{l=1}^{\infty}\sum_{m=-l}^{l}\left[\frac{(l+1)a_{lm}}
{\hat{r}^{l+2}_{tr}}-\frac{2T_{lm}(\hat{r}_{tr})}{\hat{r}_{tr}}\right]Y_{lm}(\theta,\phi)~,
\label{match4}\\ 
\frac{1}{2}\psi^{(int)}_2(\hat{r}_{tr},\theta,\phi)&=G_1(\theta,\phi)=&\frac{1}{2}
\left[ \alpha_2-\frac{\lambda_2}{\hat{r}_{tr}}-\sum_{l=1}^{\infty}
\sum_{m=-l}^{l}\frac{b_{lm}}{\hat{r}_{tr}^{l+1}}Y_{lm}(\theta,\phi)\right]~,
\label{match5}
\end{eqnarray}
where the equalities on the left-hand side arise from
the matching of the pair $(\psi^{(int)},\psi^{(lay)})$ and
identify the free angular functions (\ref{F0})-(\ref{G1}), while
those on the right-hand side arise from the matching of the pair
$(\psi^{(ext)}, \psi^{(lay)})$. We also note that (\ref{match0})
is consistent with the definition of the truncation radius and
that (\ref{match2}) is equivalent to (\ref{match1}), because from
Eq.~(\ref{ord0}) we have:
${\psi_0^{(int)}}^{''}(\hat{r}_{tr})=-(2/\hat{r}_{tr}){\psi_0^{(int)}}^{'}
(\hat{r}_{tr})$.

The free constants (\ref{alpha0})-(\ref{lambda1}) are thus easily
determined. For a given harmonic $(l,m)$ of Eqs.~(\ref{match3})
and (\ref{match4}), with $l\ge 1$, the constants $A_{lm}$ and
$a_{lm}$ are governed by the linear system with $i,j=1,2$:
\begin{equation}\label{system}
M_{ij} u_j= v_i~,
\end{equation}
where the matrix $M$ is given by
\begin{equation}
M = \left(
\begin{array}{cc}
\gamma_l(\hat{r}_{tr})&\hat{r}_{tr}^{-(l+1)}\\
-\gamma_l'(\hat{r}_{tr})\hat{r}_{tr}&(l+1)\hat{r}_{tr}^{-(l+1)}
\end{array}
\right)~,
\end{equation}
and the vectors are defined as $(u_1,u_2)=(A_{lm},a_{lm})$ and
$(v_1,v_2)=-T_{lm}(\hat{r}_{tr})(1,-2)$. Such linear system
(\ref{system}) is well posed, i.e. $det M \ne 0$.
 
To show this, we may integrate Eq~(\ref{ord1lm}) for the regular
solution, under the conditions $\gamma_l(0)=\gamma'_l(0)=0$:
\begin{equation}
\gamma_l'(\hat{r})\hat{r}^2=\int_0^{\hat{r}}[l(l+1)-R_1(\hat{r};\Psi)\hat{r}^2]\gamma_l(\hat{r})d
\hat{r}~.
\end{equation}
In the vicinity of $\hat{r}=0$ the quantity $R_1(\hat{r};\Psi)
\hat{r}^2$ is vanishingly small, so that, if the quantity
$R_1(\hat{r};\Psi)\hat{r}^2$ remains smaller than $l(l+1)$, then
the regular solution, starting positive and monotonic, remains a
positive and monotonically increasing function of $\hat{r}$.
Indeed, we have checked that this condition occurs for $l \ge 2$,
because for $\Psi \in [0.5,10]$ the quantity
$R_1(\hat{r};\Psi)\hat{r}^2$ has a maximum value in the range
[4.229, 3.326]. Under this condition the function $\mu_l(\hat{r})
\equiv \gamma_l'(\hat{r}) \hat{r} +(l+1)\gamma_l(\hat{r})$ cannot
change sign, so that $det M =
\hat{r}_{tr}^{-(l+1)}\mu_l(\hat{r}_{tr})$ cannot vanish. This
argument does not work for the case $l=1$, in which the function
$\mu_l(\hat{r})$ does change sign at a point
$0<\hat{r}_0<\hat{r}_{tr}$, but we have checked directly that the
property $det M \neq 0$ is satisfied also in this case.
 
Then expressions (\ref{A2m}) and (\ref{a2m}) are easily recovered
and we conclude that for the harmonics that are not ``driven''by
the tidal potential $T({\bf \hat{r}})$ the related $a_{lm}$ and
$A_{lm}$ must vanish.
 
Similarly, for a given harmonic $(l,m)$ with $l\ge 1$,
Eq.~(\ref{F2}), which is obtained from the
second order matching, and Eq.~(\ref{match5}) can be cast in the form of
Eq.~(\ref{system}), with $(u_1,u_2)=(B_{lm},b_{lm})$ and
$(v_1,v_2)=(-g_{lm} (\hat{r}_{tr}),\hat{r}_{tr}
g'_{lm}(\hat{r}_{tr}))$. Therefore, the argument provided above
applies and we can conclude that for those harmonics for which the
particular solutions $g_{lm}(\hat{r})$ are absent (or,
equivalently, Eq.~[\ref{ord2lm}] is homogeneous), the constants
$B_{lm}$ and $b_{lm}$ must vanish. A linear system equivalent to
(\ref{system}) can be written for a fixed harmonic $(l,m)$ with
$l\ge 1$ of the solution of general order $k$ and the same
argument applies. Therefore, we conclude that the $k$-th order
term of the solution contributes only to those harmonics for which
the particular solutions are present.

\subsection{The structure of $k$-th order term}

Because we have noted (see argument introduced about
the system [\ref{system}]) that the $k$-th order term
of the solution has non-vanishing contribution only in
correspondence of those harmonics for which the component of
Eq.~(\ref{gen}) is non-trivial, the discussion about the structure
of the term reduces to the analysis of the structure of the
expansion in spherical harmonics of the right-hand side of that
equation. Recalling that the harmonic expansion of the product of
two spherical harmonics $(l_1,m_1)$ and $(l_2,m_2)$ can be
expressed by means of 3-j Wigner symbols \citetext{see \citealp{Edm60},
Eq.~[4.6.5]}, we note that the composed harmonic $(l,m)$ must
satisfy the following {\em selection rules}: (i) $|l_1-l_2| \le l
\le l_1+l_2$ (``triangular inequality"), (ii) $m_1+m_2=m$, and
(iii) $l_1+l_2+l$ must be even. The last condition holds because
in the cited expression the composed harmonic appears multiplied
by the special case of the Wigner symbol with $(l_1,l_2,l)$ as
first row and $(0,0,0)$ as second row. Bearing in mind that the
first order term is characterized by harmonics with $l=0,2$ and
corresponding positive and even values of $m$ and that the
structure of the right-hand side of Eq.~(\ref{gen}) can be
interpreted by means of the partitions of the integer $k$, it can
be proved {\em by induction} that the $k$-th order term is
characterized by harmonics with $l=0,2,...,2k$ and corresponding
positive and even values of $m$.

\section{Extension to the presence of internal rotation}
 
It is well known that in the presence of finite total angular
momentum of the system, relaxation leads to solid-body rotation
(e.g., see Landau \& Lifchitz 1967). If we denote by $\omega$ the angular
velocity of such rigid rotation and assume that it takes place
around the $z$ axis, in the statistical mechanical argument that
leads to the derivation of the Maxwell-Boltzmann distribution one
finds that in the final distribution function the single particle
energy $E$ is replaced by the quantity $E - \omega J_z$. Following
this picture, we may consider the extension of King models to the
case of internal rigid rotation. This extension is conceptually
simpler than that addressed in the main text of this paper,
because the perturbation associated with internal rotation, while
breaking spherical symmetry, preserves axial symmetry. We note
that the models described below differ from those studied by
\citet{KorAna71}, which were characterized by a different,
discontinuous truncation, and those by \citet{PreTom70}
and by \citet{Wil75}, which were characterized by a different
truncation prescription and by differential rotation.
 
The relevant physical model is that of a rigidly rotating isolated
globular cluster characterized by angular velocity
$\mbox{\boldmath$\omega$}=\omega \hat{e}_z$, with respect to a
frame of reference with the origin in the center of mass of the
cluster. We then introduce a second frame of reference,
co-rotating with the cluster, in which the position vector is
given by ${\bf r}=(x,y,z)$. In such rotating frame, the Lagrangian
describing the motion of a star belonging to the cluster is given
by:
\begin{equation}
\mathcal{L}=\frac{1}{2}(\dot{x}^2+\dot{y}^2+\dot{z}^2+2\omega\dot{y}x-
2\omega\dot{x}y)-\Phi_{cen}(x,y)-\Phi_C(x,y,z)~,
\end{equation}
\noindent where $\Phi_{cen}(x,y)=-(x^2+y^2)\omega^2/2$ is the
centrifugal potential; the energy integral of the motion (called
the Jacobi integral) is:
\begin{equation}
H=\frac{1}{2}(\dot{x}^2+\dot{y}^2+\dot{z}^2)+\Phi_{cen}+\Phi_C~.
\end{equation}
As in the tidal case, the extension of King models is performed by
considering the distribution function $f_K(H)$, as in (\ref{fK}),
for $H \le H_0$ and $f_K(H)=0$ otherwise, with $H_0$ the cut-off
constant. The dimensionless energy is defined by:
\begin{equation}
\psi({\bf r})=a\{H_0-[\Phi_C({\bf r})+\Phi_{cen}(x,y)]\}~,
\end{equation}
\noindent and the boundary of the cluster, implicitly defined as
$\psi({\bf r})=0$, is an equipotential surface for the total
potential $\Phi_C+\Phi_{cen}$. Its geometry, reflecting the
properties of the centrifugal potential, is characterized by
symmetry with respect to the $z$-axis and reflection symmetry with
respect to the equatorial plane $(x,y)$. The constant $\psi$
family of surfaces, much like the Hill surfaces of the tidal case,
is characterized by a {\em critical} surface which distinguishes
the closed from the opened ones and in which the points on the
equatorial plane are all saddle points. In these points the
centrifugal force balances the self-gravity of the cluster; their
distance from the origin, which we call {\em break-off radius}
($r_B$), can be determined from the condition:
\begin{equation}\label{breakoffradius}
\frac{\partial \psi}{\partial x}(r_B,0,0)=\frac{\partial
\psi}{\partial y} (0,r_B,0)=0~.
\end{equation}
Following the argument that we gave in Sect.~2.2 for the
tidal radius, we find a zero-th order approximation for the
break-off radius: $r_B^{(0)}=(GM/\omega^2)^{1/3}$, where $M$ is
the total mass of the cluster. The discussion about the parameter
space that characterizes these models is equivalent to the one
presented in Sect.~2.3, with:
\begin{equation}
\chi \equiv \frac{\omega^2}{4 \pi G \rho_0}~,
\end{equation}
a dimensionless parameter that measures the strength of
the rotation with respect to the central density of the cluster,
playing the role of $\epsilon$. Similarly, we may define an
extension parameter $\sigma=r_{tr}/r_B$ (instead of $\delta$, see
Sect.~2.3). For every value of the dimensionless central potential
well $\Psi$, a maximum value for the strength parameter,
$\chi_{cr}$, exists, corresponding to the critical condition in
which the boundary of the cluster is given by the critical
equipotential surface.
 
The relevant equations in the construction of these (fully)
self-consistent models can be expressed in dimensionless form by
means of the same rescaling of variables performed in Sect.~3.
Thus, for $\psi \ge 0$, the Poisson equation can be written as:
\begin{equation}\label{Poissonrot}
\hat{\nabla}^2
\psi=-9\left[\frac{\hat{\rho}(\psi)}{\hat{\rho}(\Psi)}-2\chi
\right]~,
\end{equation}
where $\hat{\rho}$ is defined as in (\ref{rho}). For
negative values of $\psi$ we should refer to:
\begin{equation}\label{vacuumrot}
\hat{\nabla}^2 \psi=18 \chi~,
\end{equation}
\noindent with the boundary conditions at the origin written as in
(\ref{cond1})-(\ref{cond2}) and at large radii given by $\psi+\chi
C \rightarrow aH_0$, where $C \equiv -(9/2)(\hat{x}^2+\hat{y}^2)$
is the dimensionless centrifugal potential.
 
The solution up to second order in terms of matched asymptotic
expansions presented in Sect.~4 can be adapted to this case
without effort. In fact, with respect to the calculation presented
in the main text only two differences occur: (i) wherever the
constant term $-\hat{\nabla}^2T=-9(1-\nu)$ appears, it must be
replaced here by $-\hat{\nabla} ^2C=18$ (the sign is the same in
the two cases, because $1-\nu < 0$, see Sect. 2.1), and (ii)
thanks to axisymmetry, in the angular part of the Laplacian the
derivative with respect to the toroidal angle $\phi$ can be
dropped and thus the terms of the asymptotic series
(\ref{asymserin})- (\ref{asymserext}) can be expanded by means of
Legendre polynomials\footnote{Following \citet{AbrSte}, we use 
Legendre polynomials as defined in (22.3.8), i.e. with Condon-Shortley 
phase, and normalized with respect to the relation (22.2.10). We 
remark that, although they are structurally equivalent to zonal 
spherical harmonics, the normalization is different.} instead of 
spherical harmonics. The latter property implies that the radial 
part of each term of the asymptotic series is characterized by 
only one index, $l$, i.e. we can write $\psi_{k,l}^{(\;)}$. We 
note that the differential operator that appears on the left-hand 
side of the relevant equations for the solution defined in the 
internal region is still $\mathcal{D}_l$, and thus also the functions 
$\gamma_l(\hat{r})$ can be introduced in the same way as before. 
As to the equations corresponding to (\ref{tau0})-(\ref{tau2}), 
the formal solutions of the equations in the boundary layer, the 
angular functions $F_i$ and $G_i$ (with $i=0,1,2$) now depend only 
on the poloidal angle $\theta$. About the external solution, an 
expression analogous to (\ref{extgen}) can be used, with the 
particular solution given by $\chi C$ instead of $\epsilon T$. 
The centrifugal potential contributes, as in the case of the 
tidal potential, only with monopole and quadrupole terms, 
explicitly:
\begin{eqnarray}
&&C_{0}(\hat{r})=-3\sqrt 2 \hat{r}^2 \label{C0}~,\\
&&C_{2}(\hat{r})=3\sqrt{ \frac{2}{5}}\hat{r}^2 \label{C2}~.
\end{eqnarray}
Finally, as a result of the matching of the pair $(\psi^{(int)},
\psi^{(lay)})$ up to second order, the expressions for the angular
functions (\ref{F0})-(\ref{G1}) and (\ref{F2})-(\ref{G2}) are
still applicable. In addition, from the matching of the pair
$(\psi^{(lay)},\psi^{ext})$ up to second order, we find that the
explicit expressions for the free constants follow
Eqs.~(\ref{alpha0})-(\ref{a2m}) and (\ref{alpha2})-(\ref{b4m}),
provided that we drop everywhere the index $m$ and we replace $3
T_{00}(\hat{r}_{tr})/(2\sqrt \pi)$ with $3 C_0
(\hat{r}_{tr})/\sqrt 2$ in (\ref{alpha1}) and
$T_{00}(\hat{r}_{tr})/ (\sqrt \pi \hat{r}_{tr})$ with $2
C_0(\hat{r}_{tr})/(\sqrt2 \hat{r}_{tr})$ in (\ref{lambda1}).
Obviously, the particular solutions $f_0$ and $g_{l}$ (with
$l=0,2,4$) are different from the ones obtained in the tidal case,
because the right-hand side of the relevant equations is
different. Also in this case, it can be proved {\it by induction}
that the $k$-th order term has non-vanishing contributions only
for $l=0,2,...,2k$.
 
For completeness, we record the explicit expression of the second
order equations in the internal region:
\begin{eqnarray}
&&\mathcal{D}_0\psi_{2,0}^{(int)}=-R_2(\hat{r};\Psi){1\over\sqrt{2}}\;
[(\psi_{1,0}^{(int)})^2+(\psi_{1,2}^{(int)})^2]~,\\
&&\mathcal{D}_2\psi_{2,2}^{(int)}=-R_2(\hat{r};\Psi){\sqrt{10}
\over7}\;
\left[{7\over\sqrt{5}}\psi_{1,0}^{(int)}\,\psi_{1,2}^{(int)}+(\psi_{1,2}
^{(int)})^2\right]~,\\
&&\mathcal{D}_4\psi_{2,4}^{(int)}=-R_2(\hat{r};\Psi){3 \sqrt
2\over7}\; (\psi_{1,2}^{(int)})^2~.
\end{eqnarray}
\noindent We remark that the Legendre expansion of the product of
two Legendre polynomials is straightforward, because the 3-j
Wigner symbols of interest all belong to the special case with
$(0,0,0)$ as second row.

\section{Extension of other isotropic truncated models}

The procedure developed in Sects.~3 and 4 can be applied also to
extend other isotropic truncated models, different from the King 
models, to the case of tidal distortions. Here we briefly describe 
the case of low-$n$ polytropes ($1<n<5$), which are particularly 
well suited to the purpose, because they are characterized by a 
very simple analytical expression for the density as a function of 
the potential; this class of models was also considered by \citet{Wei93}. 
In the distribution function that defines the polytropes
\citep[e.g., see][]{Ber00}, we may thus replace the single star energy 
with the Jacobi integral (see definition [\ref{Jacobi}]) and consider:
\begin{equation}\label{fP}
f_P(H)=A(H_0-H)^{n-3/2}~,
\end{equation}
for $H \le H_0$, and a vanishing distribution otherwise. Unlike the 
King family discussed in the main text, these models have no 
dimensionless parameter to measure the concentration of the stellar 
system, which depends only on the polytropic index $n$; in fact, the 
spherical fully self-consistent polytropes are characterized only by 
two physical scales, which are associated with the cut-off constant 
$H_0$ and the normalization factor $A$. Below we consider values of 
$n<5$, so that the models have finite radius.
Therefore, the relevant parameter space for the tidally distorted models 
is represented just by the tidal strength parameter $\epsilon$ (see 
definition [\ref{tidalp}]), which, for a given value of the index $n$, 
has a (maximal) critical value. The definition (\ref{delta}) for the 
extension parameter $\delta$ is still valid if by $r_{tr}$ we denote 
the radius of the unperturbed spherical configuration. The associated 
density functional is given by:
\begin{equation}
\rho(\psi)=\rho_0 \psi^n~,
\end{equation}
where the dimensionless escape energy is given by:
\begin{equation}
\psi({\bf r})=\{H_0-[\Phi_C({\bf r})+\Phi_T({\bf r})] \}\left(\frac{c_n}
{\rho_0}\right)^{1/n}~,
\end{equation}
with $c_n\equiv(2 \pi)^{3/2}\Gamma(n-1/2)A/n!$. The boundary of the 
perturbed configuration is defined by $\psi({\bf r})=0$, following the 
same arguments described in the main text. Here $\rho_0$ can be 
interpreted as the central density if we set $\psi({\bf 0})=1$.

For $\psi \ge 0$, the relevant equation for the construction of the 
self-consistent tidally distorted models is the Poisson equation, which, 
in dimensionless form, is given by:
\begin{equation}\label{Poissonpoly}
\hat{\nabla}^2 \psi=-[\psi^n-\epsilon (1-\nu)]~,
\end{equation}
while for negative values of $\psi$ we must refer to Eq.~(\ref{vacuum}).
Here the rescaling of variables has been performed by means of
the scale length $\zeta\equiv \sqrt{\rho_0^{1/n-1}/(4 \pi G c_n^{1/n})}$.
The relevant boundary conditions are given by $\psi({\bf 0})=1$ instead
of (\ref{cond1}), while (\ref{cond2}) and (\ref{cond3}) hold unchanged.

If the polytropic index is in the range $1 < n < 5$, the solution up to 
second order presented in Sect.~4 is fully applicable, provided that we 
note that the problem for the zero-th order term of the series (\ref{asymserin}) 
is now given by the Lane-Emden equation \citep[see, e.g.][]{Cha39}:
\begin{equation}\label{LaneEmden}
{\psi_0^{(int)}}^{''}+\frac{2}{\hat{r}}{\psi_0^{(int)}}'=-\left(\psi_0^
{(int)}\right)^n~,
\end{equation}
with $\psi_0^{(int)}(0)=1$ and ${\psi_0^{(int)}}'(0)= 0$, where the symbol
$'$ denotes derivative with respect to the argument $\hat{r}$; explicitly,
the truncation radius $\hat{r}_{tr}$ is now defined by $\psi_0^{(int)}
(\hat{r}_{tr})=0$, i.e. it represents the radius of the so-called
{\em Emden sphere}. Correspondingly, the quantities called $R_j$
in the main text must be re-defined as:
\begin{equation}
R_j(\hat{r};n) \equiv \frac{d^j\psi^n}{d\psi^j} \biggl\vert_{\psi_0^{(int)}}~;
\end{equation}
the value of $j$ at which the quantity $R_j$ may start to diverge depends 
on the index $n$. Obviously, in Eq.~(\ref{Poislay}), i.e. in the Poisson 
equation defined in the boundary layer, $\hat{\rho}(\epsilon \tau)$ must 
be replaced by $(\epsilon \tau)^n$. This makes it clear that the value of 
the polytropic index $n$ directly affects the order, with respect to the 
perturbation parameter, at which the density contribution on the right-hand 
side of Eq.~(\ref{Poislay}) comes into play and therefore changes the matching 
procedure. If $n >1$, the density contribution emerges only after the second 
order and thus the full procedure described in Sect.~4 is valid. In contrast, 
if $n \le 1$ the procedure described in the main text is applicable only up to
first order while the calculation of second order terms would require a 
re-definition of the boundary layer (as it happens for the case discussed in 
the main text when terms of order $k >3$ are desired).
In closing, we note that the procedure presented in this paper can be applied 
also to isotropic truncated models with more complicated expressions for the 
density functional (e.g. the family of models $f_{1n}$ proposed by \citet{Dav77}, 
without boundary conditions on tangential velocity, for which the density 
functionals are expressed in terms of the error function and of the Dawson 
integral), bearing in mind the last {\em caveat} about the possibility that 
the density contribution may affect at some order the boundary layer, thus 
requiring a reformulation of the results presented in Sect.~4.

\clearpage

\begin{figure}
\epsscale{0.5}
\plotone{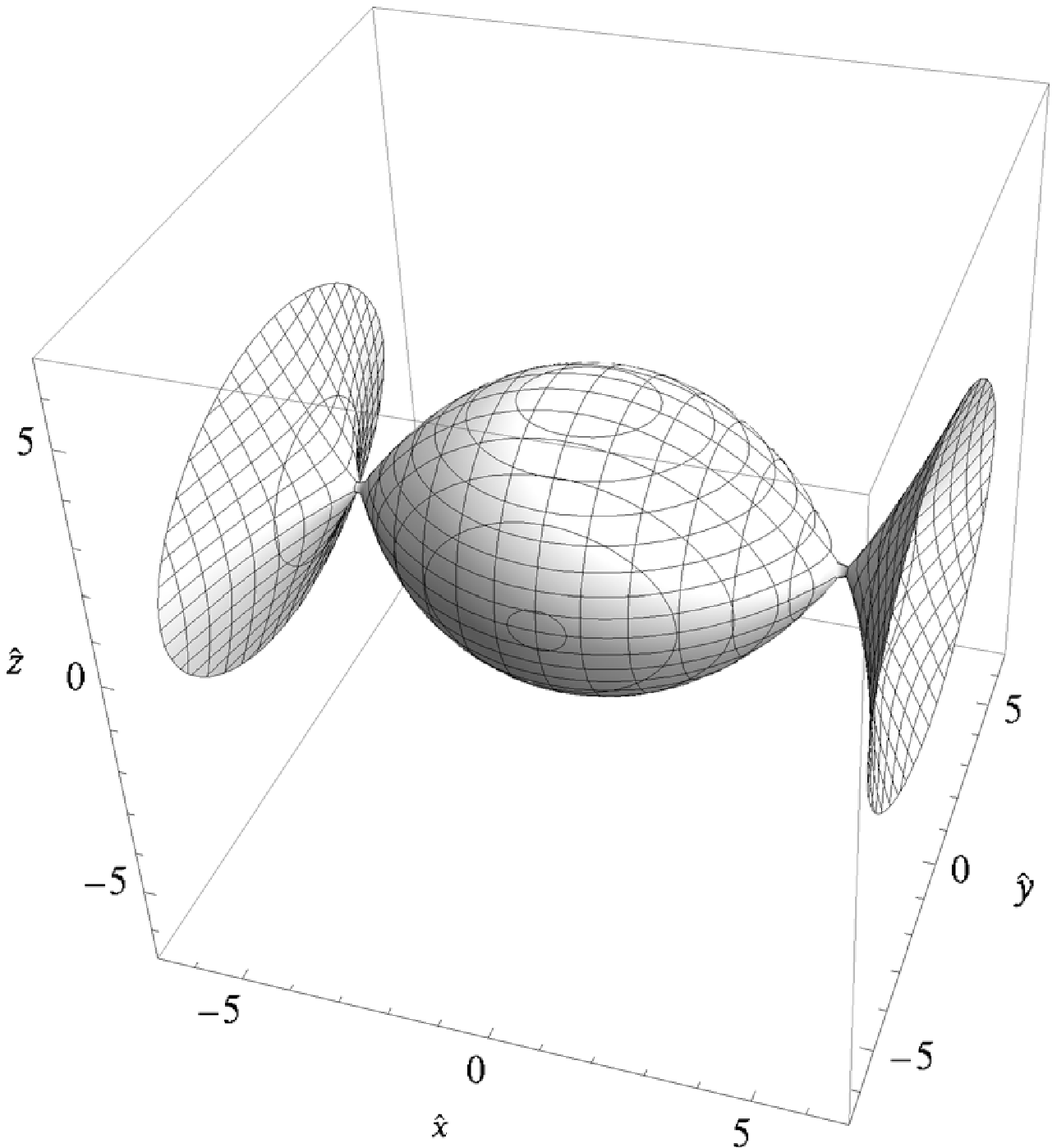}
\caption{The critical Hill surface (in dimensionless variables) for a 
second-order model with $\Psi=2$ and $\epsilon =7.043\times 10^{-4}$ 
(corresponding to $\delta_{cr}^{(2)}=0.671$); the galactic potential 
is Keplerian ($\nu=3$).\label{fig1}}
\end{figure}
 
\begin{figure}
\epsscale{1.00}
\plotone{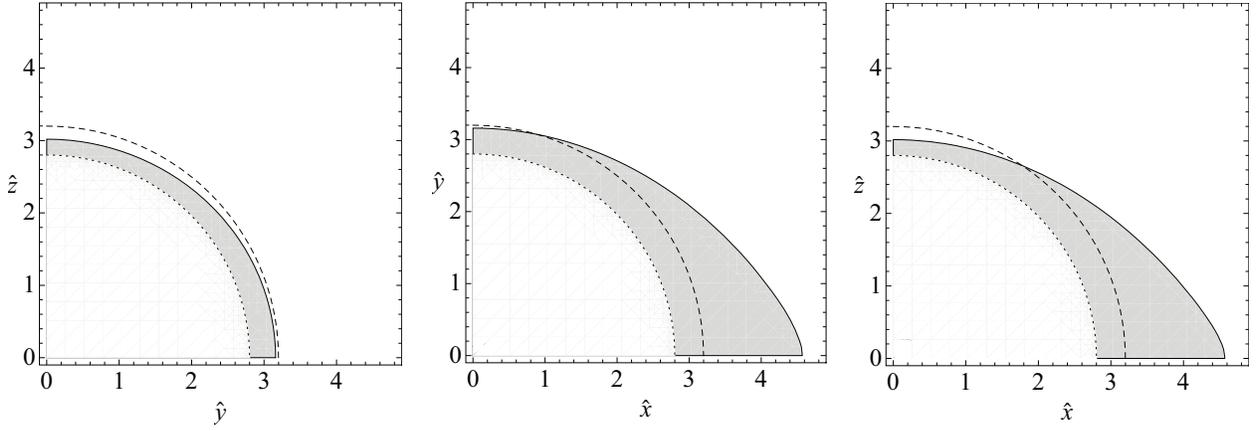}
\caption{Sections in the three coordinate planes for a second-order 
model with $\Psi=2$ and $\epsilon =7.000\times 10^{-4}$, characterized 
by high tidal distortion ($\delta=0.669\approx \delta_{cr}^{(2)}$, see 
Fig. \ref{fig1}) illustrating the boundary surface of the triaxial model (solid), 
of the internal region (dotted), and of the corresponding spherical King 
model (dashed); the filled area represents the inner region of the boundary 
layer. Note the compression and the elongation with respect to the unperturbed 
configuration in the $\hat{z}$- and $\hat{x}$-direction, respectively. The 
galactic potential is Keplerian ($\nu=3$).}
\label{fig2}
\end{figure}

\end{document}